\documentclass[12pt]{article}
\usepackage{amsmath}
\usepackage{amsthm}
\usepackage{amssymb,mathrsfs}
\usepackage{amsfonts}
\usepackage{graphicx}
\usepackage{enumerate}
\usepackage{natbib}
\usepackage{url} % not crucial - just used below for the URL 
\usepackage{color}
\usepackage{bm}
\usepackage{subcaption}
\usepackage{float}
\usepackage{cleveref}
\usepackage[ruled,vlined]{algorithm2e}
\usepackage{tikz, pgfplots}
\usepackage{booktabs}
\usetikzlibrary{intersections,decorations.pathreplacing,arrows.meta,calc,fit,patterns,shapes,shapes.misc,shapes.geometric,positioning,angles,quotes}
\usepgfplotslibrary{fillbetween}
\definecolor{dblue}{HTML}{0072B2}
\definecolor{dorange}{HTML}{D55E00}
\definecolor{dgreen}{rgb}{0.,0.6,0.}

\tikzset{
	hatch distance/.store in=\hatchdistance,
	hatch distance=10pt,
	hatch thickness/.store in=\hatchthickness,
	hatch thickness=2pt
}

\tikzset{
	invisible/.style={opacity=0},
	visible on/.style={alt={#1{}{invisible}}},
	alt/.code args={<#1>#2#3}{%
		\alt<#1>{\pgfkeysalso{#2}}{\pgfkeysalso{#3}} % \pgfkeysalso doesn't change the path
	},
}

\pgfdeclarepatternformonly[\hatchdistance,\hatchthickness]{flexible hatch}
{\pgfqpoint{0pt}{0pt}}
{\pgfqpoint{\hatchdistance}{\hatchdistance}}
{\pgfpoint{\hatchdistance-1pt}{\hatchdistance-1pt}}%
{
	\pgfsetcolor{\tikz@pattern@color}
	\pgfsetlinewidth{\hatchthickness}
	\pgfpathmoveto{\pgfqpoint{0pt}{0pt}}
	\pgfpathlineto{\pgfqpoint{\hatchdistance}{\hatchdistance}}
	\pgfusepath{stroke}
}
\makeatother

\newcommand{\blind}{1}

\begin{document}

\newtheorem{prop}{Proposition}
\newtheorem{assum}{Assumption}
\newtheorem{theorem}{Theorem}
\newtheorem{lemma}{Lemma}

\crefname{equation}{}{}
\crefname{theorem}{Theorem}{Theorems}
\crefname{lemma}{Lemma}{Lemmas}

\def\spacingset#1{\renewcommand{\baselinestretch}%
{#1}\small\normalsize} \spacingset{1}

\def\cov{\mbox{Cov}}
\def\var{\mbox{Var}}
\def\bmD{\bm{D}}
\def\bmZ{\bm{Z}}
\def\bmz{\bm{z}}
\def\bmd{\bm{d}}
\def\bmY{\bm{Y}}
\def\bmO{\bm{O}}
\def\calM{{\cal M}}
\def\bmV{\bm{V}}
\def\bmv{\bm{v}}
\def\bpsi{\bm{\psi}}
\def\bmeta{\bm{\eta}}
\def\D{\bm{D}}
\def\Y{\bm{Y}}
\def\red{\color{red}}
\def\X{\bm{X}}
\def\x{\bm{x}}
\def\cov{\text{Cov}}
\def\blue{\color{blue}}

%%%%%%%%%%%%%%%%%%%%%%%%%%%%%%%%%%%%%%%%%%%%%%%%%%%%%%%%%%%%%%%%%%%%%%%%%%%%%%

\if1\blind
{
  \title{\bf A Negative Correlation Strategy for Bracketing in Difference-in-Differences}
  \author{Ting Ye\footnote{University of Washington}, Luke Keele\footnote{University of Pennsylvania},  Raiden Hasegawa\footnote{Google},   and Dylan S. Small\footnote{University of Pennsylvania.}}
  \maketitle
} \fi

\if0\blind
{
  \bigskip
  \bigskip
  \bigskip
  \begin{center}
    {\LARGE\bf A Negative Correlation Strategy for Bracketing in Difference-in-Differences}
\end{center}
  \medskip
} \fi

\bigskip
\begin{abstract}
The method of difference-in-differences (DID) is widely used to study the causal effect of policy interventions in observational studies. DID employs a before and after comparison of the treated and control units to remove  bias due to time-invariant unmeasured confounders under the parallel trends assumption. Estimates from DID, however, will be biased if the outcomes for the treated and control units evolve differently in the absence of treatment, namely if the parallel trends assumption is violated. We propose a general identification strategy that leverages two groups of control units whose outcomes relative to the treated units exhibit a negative correlation, and achieves partial identification of the average treatment effect for the treated.  The identified set is of a union bounds form that involves the minimum and maximum operators, which makes the canonical bootstrap 
generally inconsistent and naive methods overly conservative. By utilizing the directional inconsistency of the bootstrap distribution, we develop a novel bootstrap method to construct confidence intervals for the identified set and parameter of interest when the identified set is of a union bounds form, and we theoretically establish the uniform asymptotic validity of the proposed method. We develop a simple falsification test and sensitivity analysis. We apply the proposed strategy for bracketing to study whether minimum wage laws affect employment levels.
\end{abstract}

\noindent%
{\it Keywords:} bootstrap, parallel trends, partial identification, sensitivity analysis, uniform inference
\vfill

\newpage
\spacingset{1.5} % DON'T change the spacing!

\section{Introduction}
The method of difference-in-differences (DID) is one of the most widely used strategies for policy evaluation in non-experimental settings. The simplest DID estimate is based on a comparison of the outcome differences for the treated units before and after adopting the treatment and the outcome differences for the control units. {In classic DID settings when the treated groups adopt the treatment at the same time}, the DID estimate can be obtained using fixed effects regression models and one can adjust for observed variables \citep[Ch.\ 5]{Angrist:2009}. The key advantage of DID is that it removes time-invariant bias from unobserved confounders. However, the DID method depends on a key assumption that the outcomes in the treated and control units are, in the absence of treatment, evolving in the same way over time. This key assumption is often referred to as the parallel trends assumption, which may not hold in many applications. 

The effects of minimum wage laws on employment is a key area of investigation in labor economics. In the United States, minimum wages are often set at the state or local level, which creates numerous opportunities for investigating the effects of these laws. For example, six cities in the U.S., including Seattle and Chicago, increased the minimum wage to \$15 over the last few years \citep{dube2019impacts}. Minimum wage studies have long relied on DID methods. Even very early studies of first minimum wage laws employed DID \citep{obenauer1915effect,lester1946shortcomings}. In this context,  the parallel trends assumption says that  absent the minimum wage laws, employment levels in the treated industries (or states) and in the control industries (or states) would have evolved in the same way over time. However, different industries (or states) react in different ways to the business cycle fluctuation of the economy \citep{berman1997industries}, and thus the parallel trends assumption is likely violated. Given these threats to validity, researchers need tools that exploit the strengths of DID, but depend on less stringent assumptions.

A growing literature has developed more robust inference strategies for DID designs. For example, \cite{Abadie:2005} and \cite{callaway2019difference} assume that the parallel trends assumption holds after conditioning on observed covariates. \cite{Athey:2006} assume a changes-in-changes model that is general but rules out classic measurement error on the outcome. \cite{Daw:2018aa} and \cite{Ryan:2018aa} focus on matching in DID analyses.   \citet{freyaldenhoven2019pre}  
propose to net out the effect of the unmeasured confounding by	utilizing a covariate that is affected by the same confounding factors as the outcome but is unaffected by the treatment.   \cite{manski2013deterrence, Manski:2017aa}  and \cite{Rambachan:2019aa} consider partial identification approaches to DID 	 when the variation in outcomes or violations of parallel trends are restricted to some known set.

In this article, we propose a general strategy for DID that addresses the concerns about heterogeneity in different units' outcome dynamics. We leverage two groups of control units whose untreated potential outcome relative to that for the treated units exhibit a negative correlation across the study period, i.e., when the relative outcome for one control group increases, the relative outcome for the other control group decreases (illustrated in Figure \ref{fig: monoeone_trend} and discussed in Section \ref{subsec:identification}). In this case, DID estimates can be constructed using these two control groups and used to bound (bracket) the average treatment effect for the treated units.  This general strategy builds on an idea in \citet{hasegawadid2018} but requires a weaker assumption, which can be motivated in a much broader context (see Section 3.1). We derive the identification assumption for this general strategy, which is shown to accommodate many  commonly adopted assumptions in the DID literature,  most noticeably the interactive fixed effects model with a single interactive fixed effect.

The identified set for our proposed bracketing method  takes a ``\emph{union bounds}'' form, namely the bounds can be expressed as the {union} of several intervals. {Inference for the identified set or the parameter of interest that belongs to this identified set is challenging because the union bounds form involves the minimum and maximum operators, which makes the canonical  bootstrap generally inconsistent \citep{Shao:1994aa, romano2008inference, Andrews:2009aa, Bugni:2010aa, Canay:2010aa}. For this reason, the percentile bootstrap can be overly conservative. The ``intersection-union'' approach of \cite{berger1996bioequivalence} can also be quite conservative. A related but different problem is inference on \emph{intersection bounds} \citep{chernozhukov2013intersection}, which falls into a broader class of problems where the identified set is defined by moment inequalities. Much progress has been made in this direction \citep{tamer2010partial, canay2017practical}; however, these methods cannot be directly used for union bounds because  the union bounds cannot generally be represented using moment inequalities.   Therefore, it is important to develop valid and informative inference methods for the union bounds.  To this end, by utilizing the directional inconsistency of the bootstrap distribution, we develop a novel and easy-to-implement bootstrap method to construct confidence intervals (CIs) for the identified set and parameters of interest, and we theoretically establish the uniform asymptotic validity of the proposed method. This new inference method for union bounds is itself an important contribution to the fast growing literature on inference for partially identified parameters. 

Our paper proceeds as follows. In Section \ref{sec:review}, we introduce notation and our causal framework, and  review the DID method. In Section \ref{sec:ndidb}, we develop the general bracketing strategy in DID. We introduce the identification assumption and derive the identification formula. Then we develop a bootstrap inference method and study its theoretical properties. We also develop a falsification test and sensitivity analysis. In Section \ref{sec: simu}, we examine the finite sample empirical performance of the proposed bootstrap inference method for union bounds in a simulation study. In Section \ref{sec: real}, we apply the proposed methods to study the employment effect of minimum wage laws.  Section \ref{sec: discussion}  concludes with a discussion.

\section{Review: Causal Effects Based on DID}
\label{sec:review}

We consider applications where data are observed for the treated and control units before and after the treated units adopt the treatment, while the control units are never treated. Suppose the treated units adopt the treatment between two time periods, which we denote as $t=1$ and $t=2$, and remain treated afterwards. We will refer to time period $t= 1$ as the pre-treatment period and time periods $t=2, \dots, T$ as the post-treatment periods. We write $D_{i}=1$ if individual $i$ belongs to the treated units, $D_{i}=0$ if individual $i$ belongs to the control units. One common data configuration is where the units are states, and the treatment is a change in state policy for one or more states. {In the case of  staggered adoption where the treatment is adopted by multiple states over time, we can group treated states  according to their treatment adoption time and consider each group separately; see Section S1.2 of the Supplement for details.}

As in \citet{Neyman:1923a} and \citet{Rubin:1974}, we define treatment effects in terms of potential outcomes. Let $Y_{it}^{(1)}$ represent the potential outcome for individual $i$ at time $t$ if being treated, let  $Y_{it}^{(0)}$ represent the potential outcome for individual $i$ at time $t$ if being 
untreated. We assume throughout that at each time $t$, the potential outcomes and the treated unit indicator $( Y_{it}^{(1)}, Y_{it}^{(0)}, D_i),  i=1,\dots, N_t$, are identically and independently distributed (i.i.d.) realizations of $( Y_t^{(1)}, Y_t^{(0)}, D)$. {Some caveats and reflections on such a random sampling assumption can be found in \cite{Manski:2017aa}.} Relatedly, there have been studies that evaluate the impact of within-cluster correlation arising from  a random unit-time specific component \citep{Bertrand:2004, Donald:2007}; see \citet[Section 6.5.3]{imbens2008recent} for a review. In this article, we instead view the unit-time specific components as fixed effects and propose to bracket these fixed effects rather than modeling them as random. As such, the unit-time specific components can be accounted for using our bracketing method without creating within-cluster correlation. More specific discussion on this point is given in Section~\ref{sec:inference} below. Depending on the treatment status, the observed outcomes can be expressed as  $Y_{it}=Y_{it}^{(0)}$ for $t\leq 1$, $Y_{it}=D_{i} Y_{it}^{(1)}+(1-D_{i})Y_{it}^{(0)}$ for $t=2, \dots, T$. The observed data $ \{Y_{i1}, D_{i} \}_{i=1,\dots, N_1}, \dots, \{Y_{iT}, D_{i} \}_{i=1,\dots, N_T}$ can be obtained from a longitudinal study of the same units over time or a repeated random sample. Hereafter, we drop the subscript $i$ to simplify the notation. 

We are interested in the average treatment effect for the treated units in the post-treatment periods, 
\[
ATT_t=E[Y_t^{(1)}-Y_t^{(0)}|D=1], \qquad t=2, \dots, T,
\] 
\noindent where the expectation is taken with respect to the distribution of $(Y_t^{(1)}, Y_t^{(0)}, D)$. Note that $E[Y_t^{(1)}|D=1]=E[Y_t|D=1]$ for $t=2, \dots, T$ can be identified from the observed data, but we never observe $Y_{t}^{(0)}$ for the treated units in post-treatment periods. One approach to causal identification is to use the method of DID under the parallel trends assumption 
\begin{eqnarray}
	E[ Y_{t}^{(0)}-Y_{1}^{(0)}|D=1]=E[ Y_{t}^{(0)}-Y_{1}^{(0)}|D=0], \qquad t=2, \dots, T, \label{eq: parallel}
\end{eqnarray}
which says that the treated units and the control units would have exhibited parallel trends in the potential outcomes in the absence of treatment.  With this assumption, we can use the control units to identify the change in the potential outcomes for the treated units had the units counterfactually not been treated. Thus, $ATT_t$  can be identified through
\begin{align}
	ATT_t&=E[Y_t^{(1)}-Y_t^{(0)}|D=1] \nonumber\\
	&=E[Y_t^{(1)}-Y_1^{(0)}|D=1]-E[Y_t^{(0)}-Y_1^{(0)}|D=1] \nonumber\\
	&=E[Y_t^{(1)}-Y_1^{(0)}|D=1]-E[Y_t^{(0)}-Y_1^{(0)}|D=0] \nonumber \\
	&=E[Y_t-Y_1|D=1]-E[Y_t-Y_1|D=0], \nonumber
\end{align}
where the third equality holds because of the parallel trends assumption in (\ref{eq: parallel}). In the simplest scenario, the DID estimator replaces the above conditional expectations with the corresponding sample averages.

\section{A General DID Bracketing Strategy}
\label{sec:ndidb}

In extant work, researchers have sought to relax the parallel trends assumption in various ways.  The DID bracketing method in \cite{hasegawadid2018} is one proposal that connects the outcome levels and outcome dynamics in the absence of treatment, such that the changes in outcome for different groups are ordered according to their outcome levels. Facilitated by the control group construction approach discussed in \cite{hasegawadid2018} in which units are designated to the lower (upper) control group if the average outcome is lower (higher) than the average outcome for the treated group in a prior-study period,  the two standard DID estimators based on these two control groups can bound the true treatment effect. The models and assumptions required by \cite{hasegawadid2018} are reviewed in detail in Section S1.1 of the Supplement.  In Section \ref{subsec:identification}, we will establish that the bracketing relationship holds much more broadly.

%If the unmeasured confounder $U$ does not satisfy Assumptions \ref{assum1}-\ref{assum3}, and previous outcome levels do not reflect the relative magnitude of changes in the outcome in the study period, the control groups constructed following \citet{hasegawadid2018} may fail to produce valid bounds on the treatment effect.

\subsection{Identification}
\label{subsec:identification}

We present our key partial identification assumption based on two control groups which we denote as $ a,b $. Let $\Delta_t (g)=E[Y_t^{(0)}-Y_{t-1}^{(0)}|G=g]$ be the expected change in potential untreated outcome for group $g$ from time $t-1$ to time $t$.  Let $ \Gamma_t (g) = E[Y_t^{(0)}|G=g]- E[Y_t^{(0)}|G=trt]$ be the expected potential untreated outcome for control group $ g $ at time $ t $ relative to the treated group. The \emph{only} identification assumption in our new proposal is
\begin{assum}(monotone trends) For $t= 2, \dots, T$,
	\label{assum4}
	\begin{align}\min \left\{ \Delta_t (a), \Delta_t (b) \right\}\leq  \Delta_t(trt)\leq   \max\left\{  \Delta_t (a), \Delta_t (b) \right\}. \label{eq: monotone trends}\end{align} 
\end{assum}

One can show that $ \Delta_t  (g) - \Delta_t  (trt) = \Gamma_t{(g)}- \Gamma_{t-1}{(g)}$. Therefore, Assumption \ref{assum4} has an equivalent formulation, which we state as Lemma \ref{lemma: 1}. 

\begin{lemma} \label{lemma: 1}
	Assumption \ref{assum4} is equivalent to the following: for $t= 2, \dots, T$,
	\begin{align}
		\left\{\Gamma_t(a)- \Gamma_{t-1} (a)\right\}  \left\{\Gamma_t(b)- \Gamma_{t-1} (b)\right\} \leq 0.\label{eq: negative correlation}
	\end{align}
\end{lemma}

What does Assumption~\ref{assum4} imply about the behavior of the two control groups relative to the treated group? According to  Assumption~\ref{assum4}, in every pair of adjacent time periods, the change in outcome for control group $ a $ and that for control group $ b $ provide bounds on the change in outcome for the treated group if it were untreated. The equivalent formulation in Lemma \ref{lemma: 1} provides another interesting perspective: the untreated potential outcome for the two control groups $ a,b $  relative to that for the treated units exhibit a negative correlation across the study period. In other words, if the relative outcome for control group $ a $ increases (decreases), the relative outcome for control group $ b $ decreases (increases). 

Figure~\ref{fig: monoeone_trend} provides an illustrative example of Assumption \ref{assum4} and the equivalent formulation as in Lemma \ref{lemma: 1}.  In the left figure, the treated group has the lowest outcome level across all three time periods. However, for every pair of adjacent time periods, the slope for the treated group (i.e., $\Delta_t (trt)$) is bounded by the slopes for the two control groups (i.e., $\Delta_t (a)$ and $\Delta_t (b)$).
Specifically, $\Delta_2 (b)> \Delta_2 (trt)> \Delta_2 (a)$, because from $ t=1 $ to $ t=2 $, the expected outcome for the control group $ b $  has a larger increase compared to the treated group and the expected outcome for the control group $ a $ decreases, and similarly  $\Delta_3 (a)> \Delta_3 (trt)>\Delta_3 (b)$,  which 
 implies that Assumption \ref{assum4} holds. The left figure is then translated into the figure on the right by plotting the untreated potential outcomes for the two control groups relative to the treated group (i.e., $ \Gamma_t (a)$ and  $ \Gamma_t (b) $). From the right figure, for every pair of adjacent time periods, the relative outcome for one control group increases and that for the other control group decreases. Specifically, $ \Gamma_2 (a)-\Gamma_1 (a)<0, \Gamma_2 (b)-\Gamma_1 (b)>0  $, so that (\ref{eq: negative correlation}) holds for $ t=2$; $ \Gamma_3 (a)-\Gamma_2 (a)>0, \Gamma_3 (b)-\Gamma_2 (b)<0  $, so that (\ref{eq: negative correlation}) holds for $ t=3$;  this implies the equivalent condition in Lemma \ref{lemma: 1} is satisfied.

\begin{figure}[t]
		\begin{subfigure}[t]{.6\textwidth}
			\resizebox{!}{!}{%
		\begin{tikzpicture}

\tikzstyle{dp}=[circle, inner sep=0pt, minimum size=7pt]

\tikzstyle{trobs}=[dp, draw=black, fill=black]
\tikzstyle{trlink}=[draw=black, thick]

\tikzstyle{trcf}=[dp, draw=gray, fill=gray]
\tikzstyle{trcflink}=[draw=gray,dashed, thick]

\tikzstyle{uc}=[dp, draw=dblue, fill=dblue]
\tikzstyle{uclink}=[draw=dblue, thick]

\tikzstyle{lc}=[dp, draw=dorange, fill=dorange]
\tikzstyle{lclink}=[draw=dorange, thick]

\draw [->, thick] (0,0) -- (0,5) {};
\draw [->, thick] (0,0) -- (6.5,0) {};
\node at (6.5,-0.2) {$t$};
\node at (-0.5,5) {$Y_t$ };
\draw[thick,-] (1,-0.1) -- (1,0.1) node[anchor=north,below=5pt] {1};
\draw[thick,-] (2,-0.1) -- (2,0.1) node[anchor=north,below=5pt] {2};
\draw[thick,-] (3,-0.1) -- (3,0.1) node[anchor=north,below=5pt] {3};

\node[trobs] (trtpre) at (1,1.5) {};
\node[trcf] (trcfa) at (2,1.5+0.3) {};
\node[trcf] (trcfb) at (3,1.5+0.3-0.2) {};
\draw [trcflink] (trtpre) -- (trcfa);
\draw [trcflink] (trcfa) -- (trcfb);

\node[trobs,pattern=flexible hatch, hatch distance=4pt, hatch thickness=1.5pt, pattern color=dblue] at (trtpre) {};
\node[trobs,pattern=flexible hatch, hatch distance=4pt, hatch thickness=0.5pt, pattern color=dorange] at (trtpre) {};
\node[uc] (ucposta) at (2,1.5+0.7) {};
\node[lc] (lcposta) at (2,1.5-0.3) {};
\draw [uclink] (trtpre) -- (ucposta);
\draw [lclink] (trtpre) -- (lcposta);

\node[trcf,pattern=flexible hatch, hatch distance=4pt, hatch thickness=1.5pt, pattern color=dblue] at (trcfa) {};
\node[trcf,pattern=flexible hatch, hatch distance=4pt, hatch thickness=0.5pt, pattern color=dorange] at (trcfa) {};
\node[uc] (ucpostb) at (3,1.5+0.3-0.6) {};
\node[lc] (lcpostb) at (3,1.5+0.3+0.4) {};
\draw [uclink] (trcfa) -- (ucpostb);
\draw [lclink] (trcfa) -- (lcpostb);

\node[uc] (upre) at (1,2.5+1.25) {};
\node[uc] (ua) at (2,2.5+1.25+0.7) {};
\node[uc] (ub) at (3,2.5+1.25+0.7-0.6) {};
\draw [uclink] (upre) -- (ua);
\draw [uclink] (ua) -- (ub);

\node[lc] (lpre) at (1,2.5+1.75) {};
\node[lc] (la) at (2,2.5+1.75-0.3) {};
\node[lc] (lb) at (3,2.5+1.75-0.3+0.4) {};
\draw [lclink] (lpre) -- (la);
\draw [lclink] (la) -- (lb);

\fill [opacity=0.75,fill=white] (0.5,3) rectangle (9,5);
\node[gray,right=0.15cm of trcfb, text width=3cm,align=center] {\footnotesize Counterfactual};
\node[gray,right=0.15cm of trcfb, yshift=-0.4cm, text width=3cm,align=center] {\footnotesize $ E(Y_t^{(0)}\mid G=trt) $};

\node[dblue,right=0.15cm of ub, text width=3cm,align=center] {\footnotesize Control group $ b $};
\node[dblue,right=0.15cm of ub, yshift=-0.4cm, text width=3cm,align=center] {\footnotesize 							$ E(Y_t\mid G=b) $};

\node at (3.25,6) { \footnotesize  Monotone Trends};
\node at (3.25,5.5) { \footnotesize $\min \{ \Delta_t (a), \Delta_t (b) \} \!\leq \! \Delta_t(trt) \!\leq\!   \max\{  \Delta_t (a), \Delta_t (b) \}$};

\node[dorange,right=0.15cm of lb, yshift=0.3cm, text width=3cm,align=center] {\footnotesize Control group $ a $};
\node[dorange,right=0.15cm of lb, yshift=-0.1cm, text width=3cm,align=center] {\footnotesize $ E(Y_t\mid G=a) $};

\node[dblue, above  = 0 cm of ucposta]{\footnotesize $ \Delta_2 (b) $};
\node[dorange, below =0 cm of lcposta]{\footnotesize $ \Delta_2 (a) $};
%\node[gray, below =0 cm of trcfa, yshift=5pt]{\footnotesize $ \Delta_2 (trt) $};
	\end{tikzpicture}}
% \caption{}
	\end{subfigure}%
{\begin{subfigure}[t]{.3\textwidth}
	\resizebox{!}{!}{%
	\begin{tikzpicture} 
\tikzstyle{dp}=[circle, inner sep=0pt, minimum size=7pt]

\tikzstyle{trobs}=[dp, draw=black, fill=black]
\tikzstyle{trlink}=[draw=black, thick]

\tikzstyle{trcf}=[dp, draw=gray, fill=gray]
\tikzstyle{trcflink}=[draw=gray,dashed, thick]

\tikzstyle{uc}=[dp, draw=dblue, fill=dblue]
\tikzstyle{uclink}=[draw=dblue, thick]

\tikzstyle{lc}=[dp, draw=dorange, fill=dorange]
\tikzstyle{lclink}=[draw=dorange, thick]

\draw [->, thick] (0,0) -- (0,5) {};
\draw [->, thick] (0,0) -- (4,0) {};
\node at (4,-0.2) {$t$};
\node at (-0.6,5) {$ \Gamma_t(g) $ };
\draw[thick,-] (1,-0.1) -- (1,0.1) node[anchor=north,below=5pt] {1};
\draw[thick,-] (2,-0.1) -- (2,0.1) node[anchor=north,below=5pt] {2};
\draw[thick,-] (3,-0.1) -- (3,0.1) node[anchor=north,below=5pt] {3};

\node[uc] (ua) at (1,2.5+1.25-1.5) {};
\node[lc] (la) at (1,2.5+1.75-1.5) {};

\node[uc] (ub) at (2,2.5+1.25+0.7-1.5-0.3) {};
\node[lc] (lb) at (2,2.5+1.75-0.3-1.5-0.3) {};
\node[uc] (uc) at (3,2.5+1.25+0.7-0.6-1.5-0.3+0.2) {};
\node[lc] (lc) at (3,2.5+1.75-0.3+0.4-1.5-0.3+0.2) {};
%\node[uc] (ud) at (3,3.5+0.3-0.2-1.25-0.7+0.6) {};
%\node[lc] (ld) at (3,3.5+0.3-0.2-1.75+0.3-0.4) {};

\draw[decoration={brace,mirror,raise=7pt},decorate,thick,color=dorange] (1,0.05) -- node[right=10pt, yshift=5pt] { {$ \Gamma_1(a) $} } (la);
\draw[decoration={brace,mirror, raise=7pt},decorate,thick,color=dblue] (1.2,0.05) -- node[right=10pt,yshift=-5pt] { {$ \Gamma_1(b) $} } (1.2,2.5+1.25-1.5);

\draw [uclink,dashed] (ua) -- (ub) -- (uc) ;
\draw [lclink,dashed] (la) -- (lb) -- (lc) ;

\node at (2,6) { \footnotesize  Negative correlation };
\node at (2,5.5) { \footnotesize between $ \Gamma_t(a) $ and $ \Gamma_t(b) $};

	\end{tikzpicture}}
	\end{subfigure}}
	\caption{Illustrations of Assumption \ref{assum4} (left) and Lemma \ref{lemma: 1} (right). Assumption  \ref{assum4}  holds because $\Delta_t (trt)$ is bounded in between by $\Delta_t (a)$ and $\Delta_t (b)$. Lemma \ref{lemma: 1} holds because $\Gamma_t(b)$  increases whenever $\Gamma_t(a)$ decreases, and vice versa.}\label{fig: monoeone_trend}
\end{figure}

Note that the models and conditions assumed in \cite{hasegawadid2018} (reviewed in Section S1.1 of the Supplement) imply Assumption \ref{assum4}, so the bracketing method in \cite{hasegawadid2018}  is a special case of our general strategy.

Assumption \ref{assum4} also accommodates many existing assumptions in the literature. For example, the standard DID method assumes the parallel trends assumption that requires the outcome dynamics for every group are the same, i.e., $\Delta_t (a)=\Delta_t (trt)=\Delta_t (b)$ for every $t$, under which Assumption~\ref{assum4} holds. Therefore, the  bracketing method is valid under the standard DID assumptions. Assumption~\ref{assum4} also relates to the parallel growth assumption \citep{mora2012treatment}, which requires that $\Delta_t(a)-\Delta_{t-1}(a)=\Delta_t(trt)-\Delta_{t-1}(trt)=\Delta_t(b)-\Delta_{t-1}(b)$  for every $t$. If we construct the control groups such that $\Delta_0 (a)\leq \Delta_0 (trt)\leq\Delta_0 (b)$, then the parallel growth assumption implies that $\Delta_t (a)\leq \Delta_t (trt)\leq\Delta_t (b)$ for every $t$, and thus also implies Assumption~\ref{assum4}.

More generally, Assumption \ref{assum4} can be motivated from the group-level interactive fixed effects model with a single interactive fixed effect:
\begin{align}
	Y_{gt}^{(0)}=  \alpha_t+ \eta_g+ \lambda_g F_t+ \epsilon_{gt},  \label{eq: ife}
\end{align}
for $ g\in \{a, b, trt\} $ and $ t=1,\dots, T $. Here, 
$ Y_{gt}^{(0)} $ is group $ g $'s untreated potential outcome at time $ t $, $ \alpha_t $ is a time fixed effect, $ (\eta_g, \lambda_g )$ are unobserved, time-invariant group characteristics, and $ F_t $ is an unobserved  time-varying 
factor, $ \epsilon_{gt} $ is a mean zero random error. Model \eqref{eq: ife} accommodates the heterogeneity in different groups' outcome dynamics in the absence of treatment due to the heterogeneous impact of a common latent trend, i.e., 
$ \lambda_g F_t $.  It also
includes the parallel trends and  parallel growth assumptions as special cases, which can be seen by setting $ F_t=0 $ and $ F_t-F_{t-1} = F_{t-1}-F_{t-2} $ for every $ t $, respectively. Based on model \eqref{eq: ife}, 
if  we find two control groups such that $ \lambda_a\leq \lambda_{trt}\leq \lambda_b $ or $ \lambda_a\geq \lambda_{trt}\geq \lambda_b $, then Assumption \ref{assum4}  holds.   Unlike 
previous work on treatment effects in interactive fixed effects
models  \citep{Bai:2009aa, xu2017generalized} that typically requires the number of time periods to go to infinity, Theorem \ref{prop: 4} achieves partial identification of the treatment effect with as few as two time periods by utilizing two control groups whose 
$ \lambda_a, \lambda_b $ bound $ \lambda_{trt} $.

Recall that we define the average treatment effect for the treated $ATT_t=E[Y_t^{(1)}-Y_t^{(0)}|G=trt]$ for $t=2,\dots, T$. For $t=2$, we can relate  the DID parameter using each of the two control groups $g\in \{a,b\}$ to $ATT_2$,
\begin{equation} \label{eq: tau}
	\begin{split}
		\tau_2(g)=& E[ Y_2-Y_1|G=trt]-E[Y_2-Y_1|G=g]=ATT_2+ \Delta_2 (trt) -\Delta_2 (g),  
	\end{split}
\end{equation}
where $\tau_2(a), \tau_2(b)$ are standard  DID parameters. For the case where $t>2$, define
\begin{equation} \label{eq: tau t}
		\tau_t(g) \!= \!E[Y_t- Y_{t-1}|G=trt]\! -\! E[Y_t-Y_{t-1}|G=g]\!=\!ATT_t-ATT_{t-1} \!+\! \Delta_t (trt) -\Delta_t (g), 
\end{equation}
where $\tau_t(a), \tau_t(b)$ are not standard  DID parameters because $t-1$ is also a post-treatment period when $t>2$. Under Assumption \ref{assum4}, it is true that for every $t$, $\min \{\Delta_t (trt) -\Delta_t (a),\Delta_t (trt) -\Delta_t (b) \}\leq 0$ and $\max \{\Delta_t (trt) -\Delta_t (a),\Delta_t (trt) -\Delta_t (b) \}\geq 0$. Therefore, when $ t=2 $,  $\tau_2(a)$ and $\tau_2(b)$ bound the $ATT_2$, i.e., $\min\{\tau_2(a), \tau_2(b)\} \leq ATT_2 \leq \max\{\tau_2(a), \tau_2(b)\}$; when $t>2$, we have $\min\{\tau_t(a), \tau_t(b)\} \leq ATT_t-ATT_{t-1} \leq \max\{\tau_t(a), \tau_t(b)\}$. 

We state the key partial identification result as follows. {In Section S1.2 of the Supplement, we provide an extension to adjust for covariates.}
\begin{theorem} \label{prop: 4}
	Under Assumption \ref{assum4}, the average treatment effect for the treated $ATT_t, t= 2, \dots, T$ can be partially identified through
	\begin{eqnarray}
		&&\sum_{s=2}^{t}\min\{\tau_s(a), \tau_s(b)\} \leq ATT_t \leq \sum_{s=2}^{t}\max\{\tau_s(a), \tau_s(b)\},  \label{eq: new bound}
	\end{eqnarray}
	where $\tau_2(a)$ and $\tau_2(b)$ are defined in (\ref{eq: tau}), $\tau_s(a)$ and $\tau_s(b)$ for $s>2$ are defined in (\ref{eq: tau t}).
\end{theorem}

The tightness of the bounds depends on the magnitude of violation of the parallel trends assumption, since the width of the bounds equals 
\[
\sum_{s=2}^{t}\big[\max\{\tau_s(a), \tau_s(b)\} -\min\{\tau_s(a), \tau_s(b)\}\big]=\sum_{s=2}^t\left|\Delta_s(b)-\Delta_s (a)\right|,
\]
where $|\cdot|$ denotes the absolute value. If the parallel trends assumption holds over the study period, i.e., $\Delta_s (a)=\Delta_s (b) $ for every $s$, the width of the bounds equals zero for every post-treatment period, and \eqref{eq: new bound} becomes  the identification formula from the standard DID. 

In fact, for the control groups $a$ and $b$, Theorem~\ref{prop: 4} holds under a weaker assumption than Assumption \ref{assum4}, that is the monotone trends assumption holds in a cumulative fashion. Specifically, for the bounds in (\ref{eq: new bound}) to be valid,  the sufficient and necessary condition is 
\begin{eqnarray}
	\sum_{s=2}^t \min \left\{ \Delta_s (a), \Delta_s (b) \right\}\leq \sum_{s=2}^t \Delta_s (trt)\leq  \sum_{s=2}^t \max\left\{  \Delta_s (a), \Delta_s (b) \right\}. \label{assump: cumulative monotone}
\end{eqnarray}
This cumulative monotone trends assumption is useful for scenarios when Assumption \ref{assum4} is subject to mild violations for brief time periods, but (\ref{assump: cumulative monotone}) still holds. 

{Lastly, several papers have noted that the usual parallel trends assumption may be sensitive to the functional form chosen for the outcome, for example, it may hold for $Y$ but not $\log(Y)$ or vice versa \citep{Athey:2006, kahn2020promise}. These concerns are applicable to Assumption \ref{assum4} as well. \cite{roth2020parallel} show that the parallel trends assumption holds for all functional forms under a parallel trends type assumption on the full distribution. It would be interesting to see if similar results can be obtained for the monotone trends assumption. We leave that to future work. 
}

\subsection{Estimation and Inference}
\label{sec:inference}

In Theorem~\ref{prop: 4}, we assume that two control groups satisfying Assumption \ref{assum4} are given. In practice, we typically must identify these two control groups from the candidate control units. In the application in Section \ref{sec: real}, the control groups are selected based on domain knowledge. {If data  prior to the study period are available, we can also select two control groups in a data-driven way; see Section S1.3 of the Supplement for details.} Once the two control groups are given, we can proceed with inference of the treatment effect.

First, it is important to discuss the implications of the i.i.d. assumption we invoked at the beginning of Section~\ref{sec:review}. Consider the standard model in a DID design
\[
Y_{gti} = \alpha_t+ \eta_g+\tau D_{gt} + \lambda_{gt} + \epsilon_{gti},
\]
where $ g $ indexes group, $ t $  indexes time, $ i $ indexes individual \citep{imbens2008recent}. Here, $ Y_{gti} $ is the outcome, $ D_{gt} $ is a indicator variable that equals one if group $ g $ is being treated at time $ t $, $  \alpha_t, \eta_g, \tau $ are unknown parameters, respectively representing time effects,  time-invariant group effects, and the treatment effect of interest, $ \lambda_{gt} $ is an unobserved group-time specific effect, and $ \epsilon_{gti} $ is an individual-level random error. \citet{Bertrand:2004} and \citet{Donald:2007} note that in some applied work, the $ \lambda_{gt} $ (i.e., time specific factors which affect the whole group) were effectively set to zero and ignored, which can severely understate the standard error of the DID estimator. \citet{Donald:2007} outline an approach which models the $  \lambda_{gt} $ as mean zero random factors.  In our approach, we  instead view $  \lambda_{gt} $ as fixed effects and propose to bracket the $ \lambda_{gt} $'s of treated and control groups rather than modeling them as random. As such, the existence of non-zero $ \lambda_{gt}$ can be accounted for using our bracketing method without creating within-group correlation.

In what follows, we introduce a novel bootstrap method to construct confidence intervals (CIs) for the partially identified parameter of interest $ATT_t$ and its identified set in (\ref{eq: new bound}) for $ t=2,\dots, T$. Differences between CIs for the identified set and for the parameter of interest within that set have been well-addressed in the prior literature; see, for instance, \citet{Imbens:2004aa} and \citet{Stoye:2009aa}. Algebra reveals that the identified set in (\ref{eq: new bound}) for $ATT_t$ can be equivalently formulated as 
\begin{eqnarray}
	\min_{g_s\in \{ a, b\}}\bigg\{ \sum_{s=2}^t \tau_s (g_s)   \bigg\} \leq ATT_t\leq  \max_{g_s\in \{ a, b\}}\bigg\{ \sum_{s=2}^t \tau_s (g_s)   \bigg\}, \label{eq: union bounds}
\end{eqnarray}
where the proof is in the Supplement. For example, when $ t=2 $, the bounding parameters are $\{ \tau_2(a), \tau_2(b)\} $, and their minimum and maximum form the bounds for $ ATT_2 $; when $ t=3 $, the bounding parameters are $  \{ \tau_2(a)+\tau_3(a), \tau_2(a)+\tau_3(b),  \tau_2(b)+\tau_3(a), \tau_2(b)+\tau_3(b)\} $, and their minimum and maximum form the bounds for $ ATT_3 $.  In general, there are $ 2^{t-1} $ bounding parameters for $ t\geq 2 $. We call such bounds ``union bounds''.

To describe the bootstrap inference procedure, we first rearrange the data as ${\bm O}_i=(Y_{i1}, Y_{i2}, \dots, Y_{iT}, R_{i1}, \dots, R_{iT}, G_i)$, $i=1, \dots, N$, which are assumed to be i.i.d. sequence of random vectors with distribution $P\in{\cal P}$,  where  $ {\cal P} $ is a family of distributions that satisfy a very weak uniform integrability condition stated in Theorem \ref{theo}, $Y_{it} $ is the outcome for individual $ i $ at time $ t $, $ R_{it} $ indicates whether $ Y_{it} $ is observed or not, which equals 1 if we observe $ Y_{it} $, equals 0 if not, and $N $ is the total number of  individuals we observe. We assume $ R_{it} $ is independent of $ (Y_{it}, G_i) $ for every $ t $, and $ P(R_{it}=1) $ and $ P(G_i=g) $ are strictly bounded away from zero for every $ t $ and $ g $. This data configuration enables the proposed inferential method to account for arbitrary serial correlation among multiple observed outcomes for an individual. Suppose based on $ \bm{O}_1, \dots, \bm{O}_N $, we compute a vector of sample means denoted by $\bar{{\bm X}}_N$, and ${\bm \mu} (P)=E_P( \bar{{\bm X}}_N)$.  Let $\{ \theta_j (P)=\theta_j({\bm \mu (P)}), j=[k]\}$ be the set of bounding parameters and let $\hat{\theta}_{Nj}=\theta_j(\bar{{\bm X}}_N )$ be their estimators, where $ k  $ is a finite number and $[k] = \{1,\dots, k\}$. The parameter of interest $ \psi_0 (P)$  belongs to the identified set $ \Psi_0(P)=[\theta_{\min}(P), \theta_{\max} (P)] $, where $\theta_{\min} (P) = \min_{j\in[k]}  \theta_j (P) $ and $\theta_{\max} (P) = \max_{j\in[k]} \theta_j (P) $.   The goal is to construct uniformly valid CIs for $ \Psi_0(P)  $ and $ \psi_0(P)$ in the asymptotic sense.

As an illustration, in the setting when the parameter of interest is $ ATT_2 $, 
\begin{displaymath}
	\bar{{\bm X}}_N= \left[
	\begin{array}{c}
		\frac{\sum_{G_i=trt} Y_{i2} R_{i2}}{\sum_{G_i=trt} R_{i2} } -\frac{\sum_{G_i=trt}  Y_{i1} R_{i1}}{\sum_{G_i=trt}  R_{i1}}\\
		\frac{\sum_{G_i=a}  Y_{i2} R_{i2}}{\sum_{G_i=a}  R_{i2}} -\frac{\sum_{G_i=a} Y_{i1} R_{i1}}{\sum_{G_i=a} R_{i1}} \\
		\frac{\sum_{G_i=b}  Y_{i2} R_{i2}}{\sum_{G_i=b}  R_{i2}} -\frac{\sum_{G_i=b} Y_{i1} R_{i1}}{\sum_{G_i=b} R_{i1}}
	\end{array}\right],
	\qquad 
	{\bm \mu} (P)= \left[
	\begin{array}{c}
		E_P(Y_2-Y_1|G=trt) \\
		E_P(Y_2-Y_1|G=a)\\
		E_P(Y_2-Y_1|G=b)
	\end{array}\right], 
\end{displaymath}
and $ \theta_1(P)= E_P(Y_2-Y_1|G=trt)-  E_P(Y_2-Y_1|G=a)$, $ \theta_2(P)= E_P(Y_2-Y_1|G=trt)-  E_P(Y_2-Y_1|G=b)$. The goal is to construct uniformly valid CIs for the identified set $ [\min(\theta_1(P), \theta_2(P)), \max(\theta_1(P), \theta_2(P))] $ and  parameter of interest $ ATT_2 $.

Suppose we obtain a nonparametric bootstrap sample ${\bm O}_1^*, \dots, {\bm O}_N^*$ that is drawn from the empirical distribution based on $ {\bm O}_1, \dots, {\bm O}_N $.  Let $ \bar{{\bm X}}_N^* $ and $\hat{\theta}_{Nj}^*=\theta_j( \bar{{\bm X}}_N^* ), j\in [k] $ be the bootstrap analogues calculated based on the bootstrap sample.  The bootstrap inference method can be implemented following Algorithm 1. According to Theorem \ref{theo}, the random interval in  \eqref{eq: CI} is a uniformly valid $ 1-\alpha $ level CI for the identified set $ \Psi_0 (P) $, and thus the parameter of interest $ \psi_0 (P) $; the random interval in \eqref{eq: CI att} is a more refined uniformly valid $ 1-\alpha $ level CI  for $ \psi_0 (P)  $ that adapts to the width of the identified set.

Next, we derive the theoretical properties of the CIs (\ref{eq: CI})-\eqref{eq: CI att}.  For $m= m_N$ a sequence of positive integers tending to infinity but satisfying $ m/N\rightarrow 0$ or $m= N$, let   $\hat\theta_{N,\min}, \hat\theta_{N,\max}, \hat\theta_{N, \min}^{*b},  \hat\theta_{N, \max}^{*b}, \hat\theta_{m,\min}, \hat\theta_{m,\max}$ be as defined in Algorithm 1,   $L_m(x)=P\{\sqrt{m}(\hat\theta_{m,\min} - \theta_{\min} (P) )\leq x\}$ be the true distribution of $\hat\theta_{m,\min}$, $\hat{L}_{N,\rm mod}(x)=P_*\{\sqrt{N}(\hat\theta_{N, \min}^{*b}-\hat\theta_{N,\rm min})\leq x\}$ be the  proposed bootstrap estimator of $ L_m(x) $, where $P_*$ is the conditional probability with respect to the random generation of bootstrap sample given the original data. Analogously, define  $R_m(x)=P\{\sqrt{m}(\hat{\theta}_{m,\max} -\theta_{\max}(P))\leq x\}$ and $\hat{R}_{N,\rm mod}(x)=P_*\{\sqrt{N}(\hat\theta_{N, \max}^{*b} -\hat\theta_{N,\max})\leq x\}$.  Theorem \ref{theo} lays the theoretical foundation for the proposed inference procedure.  
\begin{theorem} \label{theo} For $ t=1,\dots, T $ and $ g=a, b, trt $, 
	suppose that 	$r_{igt}=  \{Y_{it}- E(Y_{it}\mid G_i=g)\} I(G_i= g) R_{it} / P(G_i=g, R_{it}=1)$ is uniformly integrable in the sense that
	\[
	\lim_{\lambda\rightarrow \infty} \sup_{P\in \mathcal{P}} E_P\left\{  \frac{r_{igt}^2}{\var (r_{igt}) }  I\left( \frac{|r_{igt}|}{\var (r_{igt})^{1/2}} >\lambda\right) \right\}  =0,
	\]
	 $ \theta_j(\bm\mu) = \bm c_j^T \bm\mu $ with $\bm c_j $ a vector of fixed constants for $ j \in[k]$, and either (S1) or (S2) holds:  \\
	(S1)  $\lim_{N\rightarrow  \infty }\inf_{P\in \mathcal{P}} \inf_{j\in [k]: \theta_j (P) \neq \theta_{\max}(P)} \sqrt{N} ( \theta_{\max}(P) - \theta_j(P) )= \infty $  and  \\
	$\lim_{N\rightarrow  \infty }\inf_{P\in \mathcal{P}} \inf_{j\in [k]: \theta_j (P) \neq \theta_{\min}(P)} \sqrt{N} (  \theta_j(P) - \theta_{\min}(P)  )= \infty $. Set  $m= N$.\\
	(S2) $m \rightarrow \infty$ and $m/N\rightarrow 0$, as $N\rightarrow \infty$. 
	
	\noindent
Then, 
	%Suppose there exists a constant $\kappa>0$,  for every $j$,  $\theta_j- \min_{j'} \theta_{j'}$ either equals zero or larger than $ \kappa $, $\max_{j'} \theta_{j'}-  \theta_j$ either equals zero or larger than $ \kappa $, then  \\
	(a) $\lim_{N\rightarrow\infty} \sup_{P\in{\cal P}} \sup_{x\in {\cal R}} \{\hat{L}_{N, \rm mod}(x)-L_m(x)\}  \leq    0$ and  $\lim_{N\rightarrow\infty} \inf_{P\in{\cal P}} \inf_{x\in {\cal R}} \{\hat{R}_{N,\rm mod}(x)-R_m(x)\}  \geq    0$. \\
	(b) Let $c_L^* (p) =\inf\{ x\in{\cal R}: \hat{L}_{N, \rm mod}(x)\geq p\}$, $c_U^*(p)=\sup\{ x\in{\cal R}: \hat{R}_{N,\rm mod}(x)\leq p\}$, then 
	\begin{align}
		&\lim_{N\rightarrow\infty} \inf_{P\in{\cal P}} P\left\{ \sqrt{m}(\hat{\theta}_{m, \min} -\theta_{\min} (P))\leq c_L^* (p)\right\}\geq p, \nonumber\\
		&\lim_{N\rightarrow\infty} \inf_{P\in{\cal P}} P\left\{ \sqrt{m}(\hat{\theta}_{m, \max}- \theta_{\max}(P))\geq c_U^* (1-p)\right\}\geq p \nonumber.
	\end{align}
	(c) Set $ p=1-\alpha/2 $, then  
	\begin{align}
		CI_{1-\alpha}\equiv \left[ \hat{\theta}_{m, \min} - m^{-1/2}c_L^*(1-\alpha/2), \hat{\theta}_{m, \max} - m^{-1/2}c_U^*(\alpha/2)\right] \label{eq: CI theo}
	\end{align} is a uniformly valid $1-\alpha$ level CI for  the identified set $\Psi_0 (P)=[ \theta_{\min}(P), \theta_{\max}(P) ]$, i.e., $ 	\lim_{N\rightarrow\infty} \inf_{P\in{\cal P}} P\big( \Psi_0(P) \subseteq CI_{1-\alpha} \big) \geq 1-\alpha  $.\\
	(d) Let $ \hat{w}^{+} = \hat{w} I(\hat{w} >0) $, where $ \hat{w}= \{ \hat{\theta}_{m,\max}- m^{-1/2}c_U^*(1/2)\}-  \{ \hat{\theta}_{m, \min}- m^{-1/2}c_L^*(1/2)\}$,  and $ \hat{p}= 1- \Phi(  \rho \hat{w}^+) \alpha$, where  $ \Phi(\cdot) $ is the standard normal cumulative distribution function, $ \rho $ is a sequence of constants satisfying $ \rho\rightarrow\infty $, $ m^{-1/2} \rho\rightarrow 0 $ and $ \rho| \hat{w}^+ - (\theta_{\max}(P)-\theta_{\min}(P)) | \xrightarrow{P}0$, where $ \xrightarrow{P} 0$ denotes convergence in probability, then
	\begin{align}
		CI^{\psi}_{1-\alpha}\equiv \left[  \hat{\theta}_{m, \min}- m^{-1/2}c_L^*(\hat{p}), \hat{\theta}_{m, \max}- m^{-1/2}c_U^*(1-\hat{p})\right] 
	\end{align}
	is a uniformly valid $1-\alpha$ level CI for  the partially identified parameter $ \psi_0(P)$, i.e., {$ 	\lim_{N\rightarrow\infty} \inf_{P\in{\cal P}}\inf_{\psi_0(P)\subseteq \Psi_0(P)} P\big(\psi_0(P)\in CI^\psi_{1-\alpha} \big)\geq 1-\alpha  $.}
	
\end{theorem}

The proof is in the Supplement.  Note that the uniform integrability condition is also required for the standard percentile  bootstrap  \citep{Romano:2012aa} and is important for the uniform asymptotic validity of our bootstrap procedure. As discussed in \cite{Romano:2012aa}, CIs satisfying the uniform asymptotic validity are more desirable compared to  CIs that only satisfy the pointwise asymptotic validity. 

From Theorem \ref{theo}(a), the bootstrap estimators $ \hat{L}_{N,\rm mod}(x),  \hat{R}_{N,\rm mod}(x)$ are not consistent for the true distributions $ L_m(x) , R_m(x)$, and this is caused by the possibility that there can be  more than one bounding parameters equal to $ \theta_{\min} (P) $ and   $\theta_{\max} (P) $. Also, this inconsistency is directional, particularly, $ \hat{L}_{N,\rm mod}(x) $ tends to be smaller than $ L_m(x) $, and  $  \hat{R}_{N,\rm mod}(x)$ tends to be larger than $ R_m(x) $.  Given that the goal is to construct a CI with asymptotic coverage probability at least $ 1-\alpha $, using critical values $ c_L^*(p), c_U^*(p) $ satisfying $ \hat{L}_{N, \rm mod} (c_L^*(p))\geq p, \hat{R}_{N, \rm mod}(c_U^*(p)) \leq p$, we have asymptotically $ L_m (c_L^*(p))\geq \hat{L}_{N, \rm mod} (c_L^*(p)) \geq p $, $  R_m(c_U^*(p))\leq  \hat{R}_{N, \rm mod}(c_U^*(p)) \leq p$. This property lays the ground for the proposed bootstrap inference method. 

Theorem \ref{theo} contains two choices of $m$. When it is safe to assume the conditions in Scenario (S1) hold, i.e., uniformly over $P\in \mathcal{P}$, the bounding parameters that achieve the minimum and maximum are well-separated from the other parameters, then we can set $m= N$, which makes $d_{j, \min } = d_{j,\max} = 0$ for all $j$. Setting $m=N$ makes more efficient use of the data and tends to have shorter CIs because the widths of the CIs are proportional to $m^{-1/2}$. In this case, the proposed bootstrap procedure is still not the same as the standard percentile bootstrap CI \begin{align}
		\left[ Q_{\alpha/2} \left(\{\min_j \hat{\theta}_{Nj}^{*b} \}_{b\in[B] }\right), Q_{1-\alpha/2} \left(\{\max_j \hat{\theta}_{Nj}^{*b}\}_{b\in[B]} \right)\right]     \label{eq: perc boot}
	\end{align} 	
	that uses the $ \alpha/2 $ quantile as the lower end and $ 1-\alpha/2 $ quantile as the upper end.  The proposed bootstrap inference procedure in Algorithm 1 subtracts the $ 1-\alpha/2 $ quantile from the lower end and the $ \alpha/2 $ quantile from the upper end. As shown by the simulation in Table \ref{tb: simu}, due to the skewness of the distributions of $ \min_j \hat{\theta}_{Nj}^{*b} $ and $ \max_j \hat{\theta}_{Nj}^{*b} $, the proposed bootstrap inference procedure in Algorithm 1 not only guarantees adequate coverage probability but also  produces shorter CIs compared to the standard percentile bootstrap. When the conditions in Scenario 1 are in doubt, Theorem \ref{theo} indicates that a more conservative bootstrap procedure with $m$ satisfying the conditions in (S2) can still ensure uniform validity. In practice, we can set $m= N/\log(\log(N))$. This bootstrap procedure is motivated from \cite{guo2021inference}, but with modifications to ensure uniform validity in our setting. Note that the idea of using a subset of data to have strong validity guarantee under irregular settings has also appeared in \cite{wasserman2020universal}. {Based on extensive simulations included in Section S1.4 of the Supplement, we find that for DID bracketing, the bootstrap with $m=N$ has  adequate coverage probability even with shrinking difference between the bounding parameters that achieve the minimum (and maximum) and the other parameters. Therefore, we still recommend using the bootstrap with $m=N$ because it is more efficient.}

The interval (\ref{eq: CI}) is in fact a Monte Carlo approximation of (\ref{eq: CI theo}) in Theorem \ref{theo}(c).  Notice that the $ CI_{1-\alpha} $ in Theorem \ref{theo}(c) is also a uniformly valid CI for the parameter of interest $ \psi_0 (P)$, because the event that $ CI_{1-\alpha} $ covers the identified set $\Psi_0(P)$ implies the event that $ CI_{1-\alpha} $ covers the parameter of interest $ \psi_0(P) $. This also means that $ CI_{1-\alpha} $ as a CI for $ \psi_0 (P)$ can be further improved by taking into consideration the width of the identified set, which motivates $ CI^\psi_{1-\alpha} $ in Theorem \ref{theo}(d). 

The idea in Theorem \ref{theo}(d) is to set $ \hat{p}=1-\alpha $ when the bounds are wide enough in the sense that $ \rho( \theta_{\max} (P) - \theta_{\min}(P)) \rightarrow \lambda\in(0,\infty]$, and set $ \hat{p}=1-\alpha/2 $ if $  \rho( \theta_{\max} (P) - \theta_{\min}(P)) \rightarrow 0$,  where $ \rho $ satisfies $ \rho\rightarrow \infty, N^{-1/2} \rho\rightarrow 0 $. The use of $ \Phi(\cdot) $ is simply to smoothly connect these two ends  because $ \Phi(\rho\hat{w}^+) \in [0,1/2]$. The intuition behind this construction is that if the bounds are wide relative to the measurement error, the parameter of interest $ \psi_0 (P)$ can only be close to at most one boundary of the identified set, so  the asymptotic probability that $ \psi_0(P) $ is more extreme than the other boundary is negligible and the noncoverage risk is one-sided. This reasoning appears in \cite{Imbens:2004aa}, \cite{Stoye:2009aa} and  \cite{CLR2009}. In practice, we set $ \rho=\{m^{-1/2} \max[ c_U^* (3/4)-c_U^* (1/4), c_L^* (3/4)-c_L^* (1/4)]\}^{-1}/\log(m) $ as in Algorithm 1, which is proportional to $\sqrt{m}/\log(m)$ and is reminiscent of the parameter used in generalized moment selection for moment inequalities \citep{andrews2010inference}. A caveat of using these types of sequences of tuning parameters is that although they don't affect the asymptotic performance of the inference procedure, they may affect the finite sample performance and there is limited guidance on how to choose these parameters. There has been much effort in the moment inequality literature to produce tests that are valid in a finite-sample normal model without relying on sequences of tuning parameters, e.g., \cite{andrews2012inference} and \cite{romano2014practical}.  

We emphasize that $ CI^\psi_{1-\alpha} $ is valid uniformly with respect to the location of $ \psi_0 (P)$ in the identified set $\Psi_0(P) $ and the width of $ \Psi_0 (P)$. This uniformity is important because it ensures that the coverage probability is adequate even when  $ \psi_0(P)$ is at the boundary of $ \Psi_0 (P)$, or when the width of $ \Psi_0 (P)$ shrinks towards zero and point identification is established. In particular, the point identification scenario is very salient for the union bounds developed in Section \ref{sec:ndidb}, because when the parallel trends assumption holds over the study period, the width of the identified set in (\ref{eq: union bounds}) equals zero. Theorem \ref{theo}(d) guarantees that $ CI^\psi_{1-\alpha} $ is valid when the parallel trends assumption holds. 

{Theorem \ref{theo} also provides bias-corrected estimators for the identified set  $ \Psi_0(P)=[\theta_{\min}(P), \theta_{\max} (P)] $. In particular,  $ \hat{\theta}_{m,  \min}^{\rm med} = \hat{\theta}_{m, \min} - m^{-1/2}c_L^*(1/2)$ is a 
	half-median-unbiased estimator \citep{chernozhukov2013intersection} for $\theta_{\min}(P)$ in the sense that the lower bound estimator falls below $\theta_{\min} (P)$ with probability at least 1/2 asymptotically, i.e., 	$\lim_{N\rightarrow\infty} \inf_{P\in{\cal P}} P\{\hat{\theta}_{m,  \min}^{\rm med}\leq \theta_{\min} (P) \}\geq 1/2$. Analogously,  $ \hat{\theta}_{m, \max}^{\rm med}= \hat{\theta}_{m, \max} - m^{-1/2}c_U^*(1/2)$ is a 
	half-median-unbiased estimator for $\theta_{\max}(P)$ in the sense that the upper bound exceeds $\theta_{\max}(P)$ with probability at least 1/2 asymptotically, i.e.,   $\lim_{N\rightarrow\infty} \inf_{P\in{\cal P}} P\{ \hat{\theta}_{m, \max}^{\rm med} \geq  \theta_{\max} (P) \}\geq 1/2$.
}

Lastly,  as discussed in the introduction, the union bounds we focus on in this article are different from the intersection bounds considered in \cite{CLR2009, chernozhukov2013intersection}, in which the minimum operator appears in the upper bound and the maximum operator appears in the lower bound. Applying methods developed in \cite{CLR2009, chernozhukov2013intersection} to the union bounds tends to produce a CI with insufficient coverage probability. The proposed inference method generally applies to an identified set for a parameter that can be expressed as union bounds, i.e., the lower bound can be formulated as the minimum of a set of bounding parameters, and the upper bound can be formulated as the maximum of another set of bounding parameters. These two sets of bounding parameters can be different. This union bounds problem can also be addressed using the ``intersection-union'' approach by  \cite{berger1996bioequivalence}, which uses $\max_j (\hat\theta_{Nj} + z_{1-\alpha/2} \hat\sigma_{Nj})$ as an upper $1-\alpha/2 $ level CI for $\theta_{\max} (P)$, where $z_{1-\alpha/2}$ is the $1-\alpha/2$ quantile of the standard normal distribution and $\hat\sigma_{Nj}$ is the standard error of $\hat\theta_{Nj}$. The lower  $1-\alpha/2$ level CI is $\min_j (\hat\theta_{Nj} - z_{1-\alpha/2} \hat\sigma_{Nj})$. By Bonferroni's inequality, $\big[\min_j (\hat\theta_{Nj} - z_{1-\alpha/2} \hat\sigma_{Nj}),  \max_j (\hat\theta_{Nj} + z_{1-\alpha/2} \hat\sigma_{Nj})\big]$  is an $1-\alpha$ level CI for the identified set  $ \Psi_0 (P)$; the proof is in the Supplement. This approach can be conservative if the $\theta_j$'s are close together, but yields a uniformly valid CI under mild assumptions without any tuning parameters; see the Supplement for a proof. Another alternative approach is the bootstrap of \cite{fang2019inference} using the directional differentiability of the max/min function.

\subsection{Assessing the Validity of the Design}

To enhance the reliability of an observational study, it is useful to include additional analyses that explore whether the key assumptions appear plausible and whether conclusions are robust to violations of the key assumptions \citep{Rosenbaum:2010}. In this section, methods for a falsification test and a sensitivity analysis are developed.

\subsubsection{Falsification Testing}

In the context of DID, investigators often conduct a falsification test by testing for parallel trends in pre-treatment time periods \citep[Chapter 5]{Angrist:2009}. Our falsification test is similar in principle, where we test whether the monotone trends relationship holds in  a pair of unused adjacent time periods prior to the study period, say $t= t_1^*,  t_2^*$. 

%then it is plausible that the monotone trends relationship also holds in the study period ($t\geq 2$). 

%Recall that Assumption~\ref{assum4} requires that for every pair of adjacent time points in the study period, the changes in outcome for the two control groups bound the change in outcome for the treated group if the treated group were untreated.

The null hypothesis that the monotone trends assumption holds when $t=t_1^*, t_2^*$  is
$$ H_0: \min \left[ \Delta_{t_2^*} (a), \Delta_{t_2^*} (b) \right]\leq \Delta_{t_2^*} (trt)\leq \max \left[  \Delta_{t_2^*} (a), \Delta_{t_2^*} (b)\right],
$$
\noindent and the alternative hypothesis includes two possible scenarios when $H_0$ is not true:\\ (i) $\min \left\{ \Delta_{t_2^*} (a), \Delta_{t_2^*} (b) \right\}> \Delta_{t_2^*} (trt)$; (ii) $\max \left\{ \Delta_{t_2^*} (a), \Delta_{t_2^*} (b) \right\}< \Delta_{t_2^*}(trt)$.  Next, we define a set of simple hypotheses:
\begin{eqnarray}
	&&H_{a}^{i}: \Delta_{t_2^*} (a)-\Delta_{t_2^*} (trt)\leq 0, \nonumber\\
	&&H_{b}^{i}:  \Delta_{t_2^*} (trt)-\Delta_{t_2^*} (b)\leq 0, \nonumber\\
	&&H_{a}^{d}:  \Delta_{t_2^*} (a)-\Delta_{t_2^*} (trt)\geq 0, \nonumber\\
	&&H_{b}^{d}:  \Delta_{t_2^*} (trt)-\Delta_{t_2^*} (b)\geq 0. \nonumber
\end{eqnarray}
The null hypothesis $H_0$ can be written in a form of a composite null hypothesis, that is 
\[
H_0: (H_{a}^i\cap H_{b}^i)\cup (H_{a}^d\cap H_{b}^d).
\]
Let the p-values testing each individual hypothesis $H_{a}^{i}, H_{b}^{i}, H_{a}^{d}, H_{b}^{d}$  be  $p_{a}^i, p_{b}^i, p_{a}^d$, $p_{b}^d$.  From the definition of one-sided p-values, $ p_a^d=1-p_a^i,  p_b^d=1-p_b^i.$ Therefore, following  Bonferroni's method and \cite{berger1982multiparameter}, we reject $H_0$ if 
\[
%\max(\min(p_{a}^i, p_{b}^i),  \min(p_{a}^d, p_{b}^d))\leq \alpha/2.
\max(\min(p_{a}^i, p_{b}^i),  \min(1-p_{a}^i, 1-p_{b}^i))\leq \alpha/2.
\]
%Critically, this is a falsification test, such that failing to reject $H_0$ does not mean Assumption~\ref{assum4} is valid, but rejecting $H_0$ implies the data may be inconsistent with Assumption~\ref{assum4}. In a falsification test, we prefer larger p-values, since this is better evidence of assumption plausibility. 

{It is important to emphasize the limitations of falsification tests as they pertain to the prior-study period  whereas Assumptions 1 is for the study period. Moreover, non-rejection of the falsification tests hypotheses does not provide any evidence in favor of the identifying assumptions. At best, a non-rejection provides some assurance that the data does not outright refute the premises of the main analysis. Lastly, a number of studies have noted that DID falsification tests may have low power in finite samples and they may distort estimation and inference \citep{roth2019pre,kahn2020promise,bilinski2018seeking,hartman2018equivalence}. Similar issues arise in our setting as well. The limitations of falsification tests motivate the sensitivity analysis we outline next.
}

%However, it is worth noting that -- as with tests of pre-trends for the usual DID design -- these tests based on pre-treatment time periods may have low power in finite samples and conditioning analysis on the result of these pre-tests can distort estimation and inference; see, e.g., \cite{kahn2020promise, bilinski2018seeking} and \cite{roth2019pre}. These concerns stress the importance of conducing sensitivity analyses as discussed in Section  \ref{subsec: SA}. 

%Visual inspection of the outcome trends before the study period can also be performed by plotting average pre-treatment outcomes.  If every slope for the treated group is bounded in between by the slopes for the two control groups, or equivalently if the average outcomes for the two control groups after subtracting the outcomes for the treated group exhibit a negative correlation, it provides visual evidence that Assumption~\ref{assum4} is plausible.

\subsubsection{Sensitivity Analysis}
\label{subsec: SA}

We develop a sensitivity analysis to evaluate how sensitive the conclusion is to violations of Assumption~\ref{assum4}. A sensitivity analysis is used to quantify the degree to which a key identification assumption must be violated in order for a researcher's original conclusion to be reversed. There is a large and growing literature on sensitivity analysis, e.g., \cite{Rosenbaum:1987, Imbens:2003, ding2016sensitivity} and \cite{fogary2017avg}. 

There are two  scenarios when Assumption \ref{assum4} is violated at time $t$:  (i) $\min \left\{ \Delta_t (a), \Delta_t (b) \right\}> \Delta_t(trt)$; or (ii) $\max \left\{ \Delta_t (a), \Delta_t (b) \right\}< \Delta_t(trt)$. We use  two sensitivity parameters $\gamma_t$ and $\delta_t$ for these two scenarios at time $t$, where $\gamma_t$ is for scenario (i) and $\delta_t$ is for scenario (ii), and we introduce  the following sensitivity assumption.
\begin{assum} [Sensitivity] \label{asump: 5}
	For the given non-negative sensitivity parameters\\ $\{\gamma_t, \delta_t\}_{t\geq 2}$, the two control groups $a $ and $b$, and $t=2,\dots, T$,
	$$\min \left\{ \Delta_t (a), \Delta_t (b) \right\}-\gamma_t\leq \Delta_t(trt)\leq \max \left\{  \Delta_t (a), \Delta_t (b) \right\}+\delta_t.$$
\end{assum}
\noindent {Assumption \ref{asump: 5} states that violation to the monotone trends assumption is bounded, which is similar in principle to the bounded-variation assumption in \cite{Manski:2017aa}.} When $\gamma_t=\delta_t=0$ for every $t$,  Assumption~\ref{asump: 5} degenerates to Assumption \ref{assum4}, under which the bounds in  (\ref{eq: new bound}) are valid. Next, we derive the bounds and CI for $ATT_t$ under Assumption~\ref{asump: 5}, which will serve as a basis for the sensitivity analysis. 
\begin{theorem}\label{prop: SA} Under Assumption \ref{asump: 5}, for $ t=2,\dots, T $, \\
	(a) The treatment effect for treated $ATT_t$ can be partially identified through
	\begin{eqnarray}
		\sum_{s=2}^{t}\min\{\tau_s(a), \tau_s(b)\} - \sum_{s=2}^t \delta_s\leq ATT_t \leq \sum_{s=2}^{t}\max\{\tau_s(a), \tau_s(b)\}+\sum_{s=2}^t \gamma_s , \label{eq: bound SA}
	\end{eqnarray}
	where $\tau_s(a), \tau_s(b)$ are in (\ref{eq: tau})-(\ref{eq: tau t}), $\delta_s , \gamma_s$ are sensitivity parameters defined in Assumption \ref{asump: 5}.\\
	(b) Further assume the conditions in Theorem \ref{theo}, let  $[\hat{l}_{t}, \hat{r}_t]$ be the uniformly valid $1-\alpha$ CI for $ATT_t$ (or the identified set) developed using Theorem \ref{theo} (c)-(d) under Assumption \ref{assum4}, then $[\hat{l}_t- \sum_{s=2}^t \delta_s, \hat{r}_t+\sum_{s=2}^t \gamma_s]$  is a uniformly valid $1-\alpha$ CI  for $ATT_t$ (or the identified set) under Assumption \ref{asump: 5}.
\end{theorem}

The proof is in the Supplement. The CI $ [\hat{l}_t, \hat{r}_t] $ can be constructed using the method developed in Theorem \ref{theo}. The form of the CI under Assumption \ref{asump: 5} illustrates how large $\gamma_s, \delta_s$ have to be in order for the study conclusion to be materially altered. If $\hat{l}_{t}, \hat{r}_t$ are both positive, $\sum_{s=2}^t\delta_s=\hat{l}_t$ would suffice to explain away the treatment effect, which requires that  $\sum_{s=2}^t \Delta_s (trt) \geq \sum_{s=2}^t\max \{\Delta_2 (a), \Delta_2 (b)\}+\hat{l}_t$. If this hypothesized  scenario is unlikely to happen in practice, the observed positive treatment effect is robust. Similarly, if $\hat{l}_t, \hat{r}_t$ are both negative, $\sum_{s=2}^t\gamma_s= \hat{r}_t$  would suffice to explain away the treatment effect, which requires that  $\sum_{s=2}^t \Delta_s (trt) \leq \sum_{s=2}^t\min \{\Delta_2 (a), \Delta_2 (b)\}-\hat{r}_t$. If this hypothesized scenario is unlikely to happen in practice, the observed negative treatment effect is robust. 

\section{Simulations}
\label{sec: simu}

We empirically evaluate the performance of the proposed bootstrap inference methods for union bounds. {Here we only present results using  the recommended bootstrap method (i.e., with $m=N$) under the following two scenarios.} Additional simulations including the bootstrap with $m/N\rightarrow 0$ and under more scenarios are in Section S1.4 of the Supplement.
\\
{\bf Case I: parallel trends:} $ E[Y^{(0)}_1 |G=trt] = 3, E[Y^{(0)}_1|G=a] = 10, E[Y^{(0)}_1|G=b] = 4$, $ \Delta_t (trt)= \Delta_t (a)= \Delta_t (b)\equiv \Delta_t $ for every $ t $, where $ \Delta_2=1, \Delta_3=-2, \Delta_4=-1 $.\\
{\bf  Case II:  partially parallel trends:} $ E[Y^{(0)}_1 |G=trt] = 3, E[Y^{(0)}_1|G=a] = 10, E[Y^{(0)}_1|G=b] = 4$, $ \Delta_2 (trt)=1,  \Delta_3 (trt)=-4, \Delta_4 (trt)=1, \Delta_2 (a)=1,  \Delta_3 (a)=-1, \Delta_4 (a)=1,  \Delta_2 (b)=2,  \Delta_3 (b)=-4, \Delta_4 (b)=1  $.

In both scenarios, the simulated data resembles a longitudinal study where $ N=1000 $ individuals are followed for $ T=4 $ time points. The group indicators $ G_i $ are given, where $ P(G=trt)=0.3 $, $ P(G=a) =0.2$, $ P(G=b)=0.5 $. The average treatment effects for the treated group are $ATT_1=0, ATT_2=2, ATT_3=3, ATT_4=1 $. The observed outcomes are generated from $ Y_{it}=E(Y_t^{(0)}|G) + ATT_t +\varepsilon_{it}$,  for $ t=1,\dots, T $, where $ \varepsilon_{it}$'s are independent from the standard normal distribution. We set the number of bootstrap iterations $ B=300 $ and significance level $ \alpha=0.05 $. 

Table \ref{tb: simu} reports the simulation average of half-median unbiased estimators of the bounds. It also shows   the average length and the empirical probability that the CI covers the parameter of interest $ ATT_t$ (i.e., coverage probability) for the proposed bootstrap CIs  in  \eqref{eq: CI}-\eqref{eq: CI att}, respectively for the identified set and $ ATT_t $. We compare with the intersection-union method of \cite{berger1996bioequivalence}  formed by taking the union of  every  $ 1-\alpha $ level CI for $ \sum_{s=2}^t \tau_s (g_s),$  $ g_s\in\{a,b\}$, and the percentile bootstrap CI in \eqref{eq: perc boot}.

We summarize the key findings as follows. First, all the methods produce CIs with adequate coverage probability, indicating the validity of all four types of CIs. Second, the intersection-union and the standard percentile bootstrap methods can both be unnecessarily conservative, especially when the width of the bounds is small (Case I) and the number of the bounding parameters is large ($ t=4 $). Overall, the proposed bootstrap methods improve significantly over the two comparison methods, as they produce shorter CIs with correct coverage probability. When one is interested in a CI for the partially identified parameter itself rather than the identified set, the bootstrap CI tailored for the parameter can be tighter than that for the identified set, and the improvement is more evident when the width of the bounds is large (Case II). Lastly, the proposed bootstrap methods also provide an easy recipe to generate estimates of the bounds. In Case I where the treatment effect is point identified, the true lower and upper bounds are equal and are respectively equal to 2, 3, 1 for $t=2,3,4$, respectively. The half-median unbiased estimators $\hat\theta_{N, \min}^{\rm med}, \hat\theta_{N, \max}^{\rm med}$ are closer to the true lower and upper bounds compared to $\hat\theta_{N, \min}, \hat\theta_{N, \max}$. The result is similar but maybe to a smaller extent for Case II where the treatment effect is partially identified and the lower (upper) bounds are  equal to 1, -1, -3 (2,3,1) for $t=2,3,4$, respectively.

\vspace{-3mm}

\section{Minimum Wage Data Application}
\label{sec: real}

The Fair Labor Standards Act (FLSA) of 1938 introduced the federal minimum wage in the United States, which covered economic sectors such as manufacturing, transportation and communication, wholesale trade, finance and real estate, and affected about  54\% of the U.S. workforce. More economic sectors were included in 1961, 1966, 1974, and 1986 Amendments to the FLSA. A classification of industry by date of FLSA coverage can be found in \citet[Table A1]{derenoncourt2018minimum}. \citet{derenoncourt2018minimum} used Current Population Survey data and applied a very nice cross-industry DID design to study the effects of the 1966 FLSA. Next, we apply the proposed DID bracketing strategy to investigate the employment effect of the 1974 FLSA that extended the federal minimum wage coverage to employees in the federal government and to private household domestic service workers. 

The treated group is comprised of 797 employees that work in either private households or for the federal government -- the two industries for which the minimum wage was added by the 1974 FLSA. To select two control groups, we use work in economics on the correlations between industries and gross domestic product (GDP). According to \citet[Table 1]{berman1997industries}, employment in private households and federal government (the two industries in the treated group) tends to have weak positive correlation with GDP. Employment in construction and retail trade (two industries covered by the 1961 FLSA), however, tends to have strong positive correlation with GDP, while employment in agriculture, entertainment and recreation services, nursing homes and other professional services, and hospitals (four industries covered by the 1966 FLSA) tends to be negatively correlated with GDP. We designate the 798 employees in the first set of industries as control group $ a $, and the 1568 employees in the second set of industries as control group $ b $. Under the interactive fixed effects model in \eqref{eq: ife}, Assumption \ref{assum4} is plausible with GDP conceptualized as the time-varying latent factor $F_t$.

Following \cite{derenoncourt2018minimum}, we restrict our sample to all prime-age workers aged 25 to 55, run the cross-industry design at the industry $ \times $ state $ \times $ year level, and define the outcome variable as the log employment rate. We focus on the average effect of the 1974 FLSA on employment in the affected industries in 1975. We report both adjusted and unadjusted results.  As in the original study, we control for the average age, share of males, share of white workers, share of married persons, share of low education workers, and industry fixed effects using a linear model.   

First, we plot trends in the log employment rate. The upper panel of Figure~\ref{fig:etrends} plots the average outcomes for the treated and two control groups, and the lower panel of Figure~\ref{fig:etrends} plots the average outcomes for the two control groups after subtracting the average outcomes for the treated group. A visual inspection of the lower panel in Figure \ref{fig:etrends} shows that the relative average outcomes for the two control groups exhibit a negative correlation during 1972-1974, providing visual evidence that  Assumption \ref{assum4} is plausible. We also apply the proposed falsification test using the 1972-1973 data, which is non-overlapping with the 1974-1975 data that we will soon use for analysis. The p-value for the falsification test (unadjusted) equals 0.26,  for the falsification test (adjusted) equals 0.25. Therefore, we find no evidence that data in the prior study period are inconsistent with Assumption \ref{assum4}.

\begin{figure}[t]
	\centering
	\includegraphics[scale=0.45]{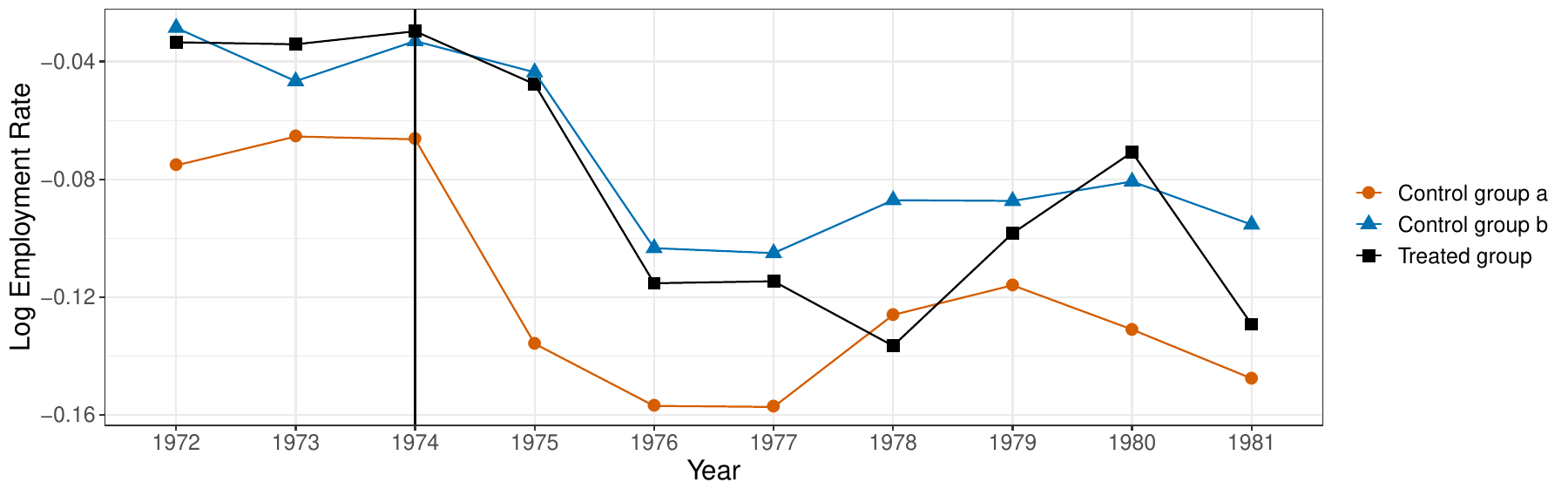}
	\includegraphics[scale=0.45]{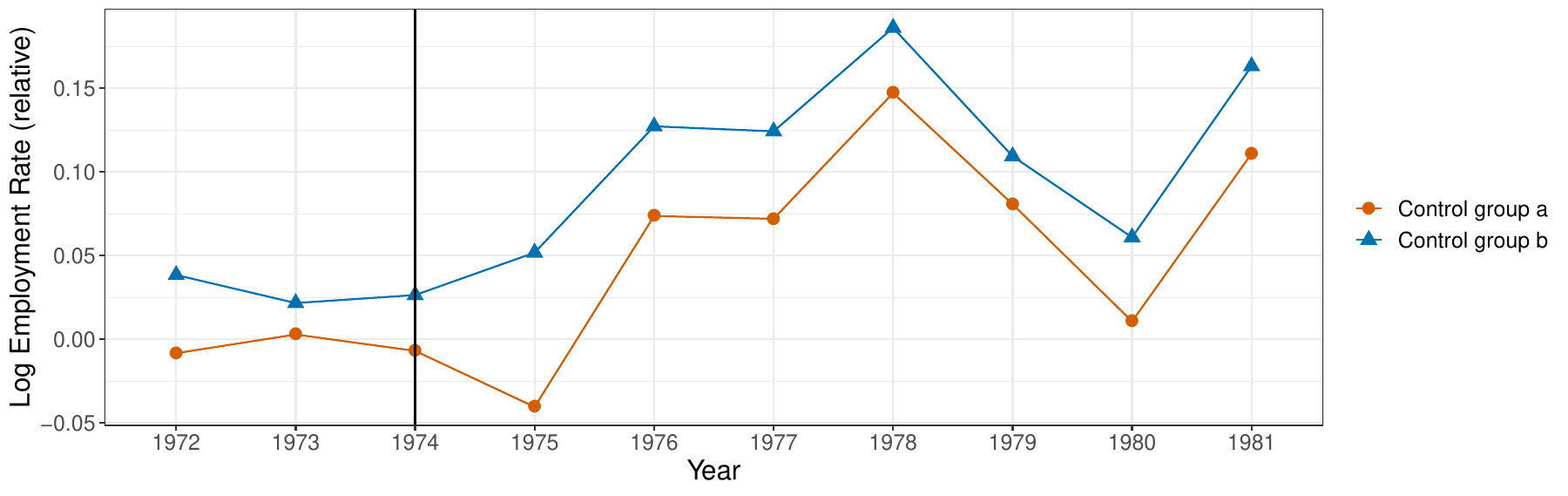}
	\caption{Upper panel:  log employment rate in control groups $ a,b $ and the treated group. Lower panel:  log employment rate  in control groups $ a,b $ relative to the treated group.} \label{fig:etrends}
\end{figure}

Table~\ref{tab.real} contains the results from the DID bracketing strategy. We also include point estimates and  95\% CIs from the standard DID method for comparison, which are directly from linear regression.   We report estimates for $ATT_t$ and CIs obtained from the proposed bootstrap method in \eqref{eq: CI}-\eqref{eq: CI att} for the identified set and $ ATT_t $ respectively, using 300 repetitions and $m=N$. In summary, DID bracketing finds no detectable effect of the 1974 FLSA on employment and this result is robust to adjusting for observed confounders. In this example, the CIs for the identified set and the parameter of interest are nearly identical, which is because the widths
of the bounds are not large compared to the measurement error and thus $ \hat p $ used in \eqref{eq: CI att} is very close to $ 1-\alpha/2 $. Based on the CIs for $ ATT_t $, we are able to rule out a negative effect on the log employment rate  more extreme than 0.029. The results based on standard DID are similar. 
Our results are also similar to the results presented in \citet[Table E1]{derenoncourt2018minimum}.

\vspace{-5mm}

\section{Concluding Remarks} 
\label{sec: discussion}
The method of difference-in-differences (DID) is widely used for policy evaluation in non-experimental settings. The DID method has the advantage of allowing flexible data structures (e.g., it works for repeated cross sectional studies or longitudinal data) and being able to remove time-invariant systematic differences between the treated and control groups. However, it is also well known that the DID method relies on a strong parallel trends assumption, which requires that in the absence of the treatment, the treated and control groups would experience the same outcome dynamics. To relax the stringent parallel trends assumption, recent work by \cite{hasegawadid2018} outlined a bracketing method, that uses two control groups with different outcome levels in the prior-study period and addresses the bias arising from a historical event interacting with the groups. 

In this work, we propose a general strategy for bracketing in DID that addresses the concerns about the heterogeneity in different units' outcome dynamics in the absence of treatment, and includes the original bracketing method in \cite{hasegawadid2018} as a special case. Critically, we leverage two control groups whose  untreated potential outcomes relative to the treated group exhibit a negative correlation, and such negative correlation naturally exists 
under the interactive fixed effects model with a single interactive fixed effect. This negative correlation appears to be reasonable in our minimum wage application, as industries fluctuate with the business cycle in different ways, some are cyclical while some are  countercyclical. {More broadly, our bracketing strategy is applicable to studies where the main concern about confounding is differential economic shocks, which includes many topics in labor economics. 
	Another general area that the method is applicable is in public health. For example, in state-level DID analyses of policies to address drug overdose deaths, a major concern is differential shocks in the availability of fentanyl. We can use states such as Ohio that were first hit by fentanyl and had high per capita overdose rates as one control group, and states such as Nebraska that were relatively insulated from  fentanyl and had low per capita overdose rates as the other control group \citep{zoorob2019fentanyl}.  Under an interactive fixed effects model with the nationwide fentanyl supply as the latent time-varying factor, these two control groups can be used to bound the potential overdose outcomes of the other states in the absence of policy interventions. 
}

%For example, it could be used
%in marketing where one might be interested in considering the effect of an advertising campaign for one product and use DID comparing the advertised product to unadvertised products.  Products which are substitutes and  products which are complements for the advertised product  might be expected to have negative correlations relative to the demand of the advertised product, enabling bracketing using negative correlations. 

Another important contribution in this work is that we develop a novel and easy-to-implement bootstrap inference method to construct uniformly valid CIs for union bounds (i.e., the identified set can be formulated as the union of several intervals). This bootstrap inference method for union bounds has the potential of being applied to broader settings.

\section*{Supplementary Material}
\vspace{-3mm}
We include the technical proofs, R code, and replication files for the minimum wage application in the Supplement.

\vspace{-3mm}
\section*{Acknowledgements}
\vspace{-3mm}

The authors would like to thank the Editor, Associate Editor and two anonymous referees for helpful comments and suggestions. The authors also thank Fredrik Savje for a very insightful discussion of the paper.

\bibliographystyle{apalike}
\bibliography{did_bib}

\begin{algorithm}[H] \label{alg}
	\SetAlgoLined
	%	\KwResult{Write here the result }
	1. Initialize $ b=1, B, \alpha, m$\;
	2. For $j\in [k]$, set $d_{j,\rm min} = (1- \sqrt{m}/\sqrt{N}) ( \hat\theta_{N, \min} - \hat\theta_{Nj} )$ and $d_{j,\rm max} = (1- \sqrt{m}/\sqrt{N}) ( \hat\theta_{N, \max} - \hat\theta_{Nj} )$, where $\hat\theta_{N,\min} = \min_{j} \hat\theta_{Nj}$ and $\hat\theta_{N,\max} = \max_{j} \hat\theta_{Nj}$\;
	3. Randomly sample a subset of data of size $m$ and calculate $\{\hat\theta_{mj}\}_{ j\in[k]},  \hat\theta_{m,\min}, \hat\theta_{m,\max}$ as their full sample analogous\; 
	4. \While{$ b\leq B $}{
		Generate a bootstrap sample of size $N$, compute $ \{\hat\theta_{Nj}^{*b}\}_{ j\in [k]} $, 
		$\hat\theta_{N,\min}^{*b}= 
		\min_j  (\hat\theta_{Nj}^{*b} + d_{j,\rm min} ),$ and $ \hat\theta_{N,\max}^{*b}=	 \max_j  (\hat\theta_{Nj}^{*b} + d_{j,\rm max} ) $\;
		Set $ b=b+1 $\;
	}
	5. Denote the $q $ sample quantile of $\{x^b\}_{ b\in [B]}$  as
	$ Q_{q} \left(\{x^b  \}_{b\in[B] }\right)$. 
	{The CI for the \emph{identified set} $\Psi_0(P)$ is}
	\begin{align}
		\begin{split}
			\left[ \hat\theta_{m,\min}- \! \sqrt{\frac{N}{m}}Q_{1-\alpha/2} \left(\{\hat\theta_{N,\min}^{*b}  - \hat\theta_{N,\min}\}_{b\in[B] }\right), \hat\theta_{m,\max} -   \! \sqrt{\frac{N}{m}} Q_{\alpha/2} \left(\{\hat\theta_{N,\max}^{*b} -  \hat\theta_{N,\max} \}_{b\in[B] } \right)\right]. \label{eq: CI} 
		\end{split}
	\end{align} 
	\!\!\!\!\! 6. {Compute $ \hat \omega =\hat\theta_{m,\max} -   \! \sqrt{\frac{N}{m}} Q_{1/2} \left(\{\hat\theta_{N,\max}^{*b} -  \hat\theta_{N,\max} \}_{b\in[B] } \right) -  \hat\theta_{m,\min}+ \! \sqrt{\frac{N}{m}}Q_{1/2} \left(\{\hat\theta_{N,\min}^{*b}  - \hat\theta_{N,\min}\}_{b\in[B] }\right)$, $\hat \omega^+ = \max (0, \hat \omega) $,  $ \rho = (\log m )^{-1} \sqrt{\frac{m}{N}}/ \max\left[Q_{3/4} \big( \{\hat\theta_{N,\max}^{*b} \}_{b\in[B] }  \big)- Q_{1/4} \big(\{\hat\theta_{N,\max}^{*b} \}_{b\in[B] }  \big) \right.,$  \\  ~~ ~ ~  $ \left. Q_{3/4} \big(\{\hat\theta_{N,\min}^{*b} \}_{b\in[B] }  \big)- Q_{1/4} \big(\{\hat\theta_{N,\min}^{*b} \}_{b\in[B] }  \big) \right]  $,
		and $ \hat p  = 1- \Phi(\rho\hat \omega^+)\alpha  $, where $ \Phi(\cdot) $ is the standard normal cumulative distribution function.}
	{The CI for the \emph{parameter of interest} $\psi_0(P)$ is}  
	\begin{align}
		\begin{split}
			\left[ \hat\theta_{m,\min}- \! \sqrt{\frac{N}{m}}Q_{\hat p} \left(\{\hat\theta_{N,\min}^{*b}  - \hat\theta_{N,\min}\}_{b\in[B] }\right), \hat\theta_{m,\max} -   \! \sqrt{\frac{N}{m}} Q_{1-\hat p} \left(\{\hat\theta_{N,\max}^{*b} -  \hat\theta_{N,\max} \}_{b\in[B] } \right)\right]. \label{eq: CI att}
		\end{split}
	\end{align} 
	\caption{Bootstrap Inference for Union Bounds}
\end{algorithm}

\begin{table}[H]
	\caption{Simulation results based on 1000 runs. The statistics reported are:   simulation average of $\hat\theta_{N, \min}$,  $\hat\theta_{N, \max}$, $\hat\theta_{N, \min}^{\rm med}$,  and  $\hat\theta_{N, \max}^{\rm med}$, the average CI length and coverage probability (CP in \%) of 95\% CIs obtained from the proposed bootstrap method in \eqref{eq: CI}-\eqref{eq: CI att} and two other methods (the intersection-union method and  standard percentile bootstrap). The sample size is $ N=1000$ and number  of bootstrap iterations is $ B=300 $. 	\label{tb: simu} } 
	\centering
	\fbox{	
		\resizebox{\textwidth}{!}{\begin{tabular}{lcccccccccccccccc}  
				&&& &   \multicolumn{6}{c}{Modified bootstrap  ($m=N$)}   & &     \multicolumn{2}{c}{Intersec.-Union }   &&\multicolumn{2}{c}{Perc. Boot.} \\
				&$\hat\theta_{N, \min}$ &   $\hat\theta_{N, \max}$ & &      $\hat\theta_{N, \min}^{\rm med}$ &   $\hat\theta_{N, \max}^{\rm med}$   &      \multicolumn{2}{c}{CI  (Set)} & \multicolumn{2}{c}{CI  ($ ATT_t $)}        &&  \multicolumn{2}{c}{CI (Set)}   &&  \multicolumn{2}{c}{CI (Set)}  \\[0.5ex] 	\cline{2-3} \cline{5-10} \cline{12-13} \cline{15-16}
				& Mean & Mean &&Mean & Mean &  Length & CP  & Length & CP   &&  Length     & CP && Length     & CP           \\  \hline 
				{\footnotesize \bf Case I  }   &                    &                    &           &       &      &         &           \\
				$ t=2 $&1.952&2.047& &1.970&2.030&0.483 & 96.7 & 0.478 & 96.7 && 0.553 & 99.0 &&0.581	&99.3\\
				$ t=3 $	&2.905 & 3.098&&2.941 &	3.063 &0.583 & 97.7 & 0.575 & 97.7 && 0.730 & 99.8 && 0.771	& 99.9\\
				$ t=4 $ &0.859&1.144&&0.913	& 1.090 &	0.672 & 98.4 & 0.661 & 98.4 && 0.893 & 100.0 && 0.955	&100.0\\
				{\footnotesize \bf Case II    }    &                    &                    &           &       &      &         &           \\
				$ t=2 $	&1.003	& 1.997&&1.003	& 1.997&1.455 & 97.9 & 1.404 & 96.8 && 1.443 & 97.7 && 1.455&	97.8 \\
				$ t=3 $		&-0.994	& 2.997&&-0.994	& 2.998&4.562 & 98.1 & 4.472 & 96.1 &&4.547	&97.8&&	4.563	&98.1 \\
				$ t=4 $ &-3.041 & 1.043&&-3.021 &	1.025	& 4.633 &	98.5 &4.542	& 96.7  &&	4.693	&98.7	&&4.728 &	99.3\\
	\end{tabular}}}
\end{table}

\begin{table}[H]
	\caption{Results from the DID bracketing and	standard DID 	for the effect of the 1974 FLSA on the log employment rate  in the affected industries in 1975.\label{tab.real}}
	\centering
	%\fbox{	
		\resizebox{0.8\linewidth}{!}{\begin{tabular}{lcccccccc}
				\toprule
				&  \multicolumn{4}{c}{DID Bracketing}                & & \multicolumn{3}{c}{Standard DID} \\
				\cline{2-4} \cline{6-9} 
				& $\hat\theta_{N,\min}^{\rm med}, \hat\theta_{N,\max}^{\rm med}$ &  CI  (Identified Set)& CI  ($ ATT_t $)&    &Mean   &   CI    \\ 
				\midrule 
				Unadjusted & -0.008, 0.050 & $ [-0.029,0.088] $& $ [-0.029,0.087]  $&& 0.012  & $ [-0.019,0.044] $ \\ 
				%1976 & $ [-0.099,0.094]  $ && -0.008 & 0.021 & $ [-0.050,0.033] $\\ 
				%1977 & $  [-0.094,0.090] $ &&  -0.007 & 0.018&$ [-0.042,0.029] $ \\ 
				Adjusted  & -0.008, 0.050 & $ [-0.031,0.081] $&  $[-0.028,0.080]  $ && 0.012 & $ [-0.014,0.038] $ \\ 
				%1976 &$  [-0.098,0.084] $ && -0.012 & 0.020 &$  [-0.051,0.027]  $ \\ 
				%1977 & $ [-0.090,0.084] ] $ &&  -0.009 & 0.017 &$  [-0.042,0.025] $\\ 
				\bottomrule
		\end{tabular}}
		%}
\end{table}

\newpage

\begin{center}
		{\sffamily\bfseries\Large
		Supplement to``A Negative Correlation Strategy for Bracketing in Difference-in-Differences''}
	
%	\noindent%
%	%EndExpansion
%	\textsf{Ting Ye\footnote{Department of Biostatistics, University of Washington}, Luke Keele\footnote{University of Pennsylvania},  Raiden Hasegawa\footnote{Google} and Dylan S. Small}\footnote{%
%		Department of Statistics and Data Science, Wharton School, University of Pennsylvania, Email: \textsf{%
%			dsmall@wharton.upenn.edu}.}
\end{center}

\pagenumbering{arabic}
\setcounter{equation}{0}
\setcounter{table}{0}
\setcounter{section}{0}
\setcounter{theorem}{0}
\renewcommand{\theequation}{S\arabic{equation}}
\renewcommand{\thetable}{S\arabic{table}}
\renewcommand{\thetheorem}{S\arabic{theorem}}
\renewcommand{\thesection}{S\arabic{section}}
\newtheorem{assumption}[theorem]{Assumption}

\section{Additional Details}

\subsection{DID Bracketing in \cite{hasegawadid2018}}

The DID bracketing method in \cite{hasegawadid2018} considers the setup with one post-treatment period (i.e., $ T=2 $), and works by partitioning the control units into two \emph{groups}--lower and upper--and uses the two standard DID estimators based on theses two control groups to bound the true treatment effect. Next, we review the proposal of \cite{hasegawadid2018} in more detail. Note that the models and assumptions in this subsection are not needed for our new proposal in Section \ref{sec:ndidb}.

Let $G$ be a group indicator, where $G=trt$ (equivalent to $D=1$) denotes the treated group, $G=lc$ denotes the lower control group and $G=uc$ denotes the upper control group. \citet{hasegawadid2018} assume the following model   that generalizes the standard DID model and changes-in-changes model \citep{Athey:2006},
\begin{eqnarray}
	Y_t^{(0)}= h(\bm U, t)+\epsilon_t,  \label{eq: model}
\end{eqnarray}
\noindent where $\bm U$ is a vector of time-invariant unmeasured confounders that may have a time-varying effect on the outcome, $ h$ is an unspecified function,  $\epsilon_t$ is an error term that captures additional sources of variation at time $t$ and $E[\epsilon_t|\bm U,G]=0$ for every $t$. Critically, $\bm U$ is time-invariant and captures the systematic difference among different groups.

In \cite{hasegawadid2018}, identification of bounds on the true treatment effect is achieved by imposing assumptions on the distribution of $\bm U$ in different groups and its effect on the outcome and the outcome dynamics in Assumptions \ref{assum1}-\ref{assum3}.
\begin{assumption}
	\label{assum1}
	The distribution of $\bm U$ within groups is stochastically ordered: $\bm U|G=lc \preccurlyeq \bm U|G=trt \preccurlyeq \bm U|G=uc$.
\end{assumption}
\noindent Note that two random vectors $A, B$ are stochastically ordered $A\preccurlyeq B$ if $E[f(A)]\leq E[f(B)]$ for all bounded non-decreasing functions $f$ \citep{hadar1969rules}.   In words, this assumption states that the unmeasured confounders are lowest in the lower control group, intermediate in the treated group, and highest in the upper control group. 

\begin{assumption}
	\label{assum2}
	The  function $h(\bm U,t)$ is bounded and increasing in $\bm U$ for every $t$. 
\end{assumption}
This assumption is natural when higher values of $\bm U$ correspond to a higher value of the outcome. The other direction is also included in this model because we can simply replace $\bm U$ with its negation. Model (\ref{eq: model}) and Assumptions \ref{assum1}-\ref{assum2} combined imply that $E[Y_t^{(0)}|G=lc]\leq E[Y_t^{(0)}|G=trt]\! \leq E[Y_t^{(0)}|G=uc]$ for every \!\! $t$.
\begin{assumption}
	\label{assum3}
	Either one of the following is satisfied: \\(a) $h(\bm U,2)-h(\bm U,1)\geq h(\bm U',2)-h(\bm U', 1)$ for all $\bm U\geq \bm U'$, $\bm U, \bm U'\in \mbox{supp} (\bm U)$; \\
	(b)  $h(\bm U,2)-h(\bm U,1)\leq h(\bm U',2)-h(\bm U', 1)$ for all $\bm U\geq \bm U'$, $\bm U, \bm U'\in \mbox{supp} (\bm U)$.
\end{assumption}
\noindent In words, the bounded function $h(\bm U, 2)-h(\bm U,1)$ is either non-decreasing in $\bm U$  over the whole support of $\bm U$, or non-increasing in $\bm U$ over the whole support of $\bm U$. Combining Assumptions \ref{assum1}, \ref{assum3} and the boundedness of $ h(\bm U,t) $, we have that 
$E[h(\bm U,2)-h(\bm U,1)|G=lc]\leq E[h(\bm U,2)-h(\bm U,1)|G=trt]\leq E[h(\bm U,2)-h(\bm U,1)|G=uc]$  or the reverse direction. 

Next, we define  $\tau_2(uc), \tau_2(lc)$, which are respectively the DID parameter using the upper control group and lower control group 
\begin{eqnarray}
	&\tau_2(uc)=E[ Y_2-Y_1|G=trt]-E[Y_2-Y_1|G=uc], \label{eq: beta1}\\
	&\tau_2 (lc)=E[ Y_2-Y_1|G=trt]-E[Y_2-Y_1|G=lc],  \label{eq: beta2}
\end{eqnarray}
Under model (\ref{eq: model}), we can relate  $\tau_2(uc), \tau_2(lc)$ with the treatment effect of interest  $ATT_2$:
\begin{eqnarray}
	\tau_2(g)&=&ATT_2+ E[Y^{(0)}_2-Y_1^{(0)}|G=trt]-E[Y^{(0)}_2-Y_1^{(0)}|G=g] \nonumber\\
	&=&ATT_2+ E[h(\bm U,2)-h(\bm U,1)|G=trt] -E[h(\bm U,2)-h(\bm U,1)|G=g],\nonumber
\end{eqnarray}
for $g\in \{uc, lc\}$. 
Additionally under Assumptions~\ref{assum1}--\ref{assum3}, one of the two DID parameters is too large and the other is too small such that we can bound $ATT_2$:
\begin{eqnarray}
	\min\{\tau_2(uc), \tau_2(lc)\} \leq ATT_2 \leq \max\{\tau_2(uc),\tau_2(lc)\} \label{eq: original did bounds}.
\end{eqnarray}
The  DID parameters $ \tau_2(uc) , \tau_2(lc)$ are identifiable from the observed data. For example, one can simply replace the conditional expectations in (\ref{eq: beta1})-(\ref{eq: beta2}) with sample analogues to obtain the corresponding DID estimators. As such, the DID bracketing method accounts for the bias arising from the time-varying effect of the unmeasured confounders $ \bm U $ by leveraging  the connections between the effect of $\bm U$ on the outcome and the effect of $ \bm U $ on the outcome dynamics using two control groups. Facilitated by the control group construction approach discussed in \cite{hasegawadid2018} in which units are designated to the lower (upper) control group if the average outcome is lower (higher) than the average outcome for the treated group in a prior-study period,  $  ATT_2 $ is partially identified via (\ref{eq: original did bounds}).

It is easy to see that the model \eqref{eq: model} and Assumptions \ref{assum1}-\ref{assum3}  imposed in \cite{hasegawadid2018} imply our Assumption \ref{assum4}, with the two control groups being control group $a,b$ and two time points.  Moreover, our general DID bracketing strategy no longer requires model (\ref{eq: model}) or Assumptions \ref{assum1}-\ref{assum3} because we now explicitly impose assumptions on how the outcomes change. This greatly widens the scope of the DID bracketing method. 

\subsection{Extensions to adjust for  covariates and staggered adoption}
Suppose that we have observed baseline covariates $X_i$ for individual $i$.  Define $\Delta_t(g,X)= E(Y_t^{(0)} - Y_{t-1}^{(0)} \mid X, G=g)$, we propose a conditional monotone trends assumption: 
\begin{assumption}\label{assump: monotone trends, X}
	For $t=2,\dots, T$, 
	\begin{align*}
		&	 \min 	\big\{ E\left[ \Delta_t(a, X)   \mid G=trt\right] ,  E\left[ \Delta_t (b, X)  \mid G=trt \right] \big\}   \\
		&\leq  E\left[ \Delta_t(trt, X)   \mid G=trt\right]  \leq \\
		&	\max	\big\{ E\left[ \Delta_t(a, X)   \mid G=trt\right] ,  E\left[ \Delta_t (b, X)  \mid G=trt \right] \big\}
	\end{align*}
\end{assumption}
Then let $\Gamma_t(g,X)= E[Y_t^{(0)}  \mid G=g, X] - E[Y_t^{(0)}  \mid G=trt, X] $. We can also show that $\Delta_t(g, X)  - \Delta_t(trt, X) = \Gamma_t(g, X)  - \Gamma_{t-1}(g, X)$. Therefore, Assumption \ref{assump: monotone trends, X} can be equivalently formulated as that for $t=2,\dots, T$, 
\[
E  [ \Gamma_t(a, X)  - \Gamma_{t-1}(a, X) \mid G=trt ]   E  [ \Gamma_t(b, X)  - \Gamma_{t-1}(b, X) \mid G=trt ]  \leq 0. 
\]

Note that the conditional parallel trends assumption required by the classical DID method says that $\Delta_t(a,X)  = \Delta_t(trt,X) = \Delta_t(b,X) $ for every $t$ under which Assumption \ref{assump: monotone trends, X} holds. Therefore, the bracketing method still requires weaker assumption compared to the classical DID in the presence of covariates. Assumption \ref{assump: monotone trends, X} can also be motivated from the interactive fixed effect model with a single interactive fixed effect for unit $i$ in group $g$ at time $t$:
\begin{align}
	Y_{git}^{(0)}  & = \alpha_t + \eta_g +\lambda_g F_t  +  \gamma_{t}^T X_{gi}   +\epsilon_{git} \nonumber \\
	&: =\eta_{gt} +   \gamma_{t}^T X_{gi}   +\epsilon_{git}  \label{eq: inter}
\end{align}
where $\eta_{gt} = \alpha_t + \eta_g + \lambda_g F_t$, for  $g\in \{ a, b,trt\}$ and $t=1,\dots, T$. Based on \eqref{eq: inter}, if we find two control groups such that $\lambda_a\leq \lambda_{trt}\leq \lambda_b$ or $\lambda_b\leq \lambda_{trt}\leq \lambda_a$, then Assumption \ref{assump: monotone trends, X} holds.

The target parameter remains the average treatment effect for the treated $ATT_t = E[ Y_t^{(1)} -Y_t^{(0)} \mid G=trt]$ for $t=2,\dots, T$. For each $t$, we can calculate the DID parameter using each of the two control groups by outcome regression (OR), inverse probability weighting (IPW), and doubly robust (DR) methods as in \cite{callaway2019difference} as follows,
\begin{align}
	\tau_{t,\rm ipw} (g)  & = E\left[ \left\{  \frac{I(G=trt)}{P(G=trt)}  - \frac{ \frac{p_{g} (X)I(G=g)}{1-p_{g} (X)  }}{ E\left[ \frac{p_{g} (X)I(G=g)}{1-p_{g} (X)  }\right]}  \right\} (Y_t - Y_{t-1}) \right], \label{eq: ipw}\\
	\tau_{t,\rm  or} (g)  & = E\left[  \frac{I(G=trt)}{P(G=trt)}  \{ Y_t - Y_{t-1} - m_{g, t} (X) \} \right],  \label{eq: or}\\
	\tau_{t,\rm  dr} (g)  & = E\left[ \left\{  \frac{I(G=trt)}{P(G=trt)}  - \frac{ \frac{p_{g} (X)I(G=g)}{1-p_{g} (X)  }}{ E\left[ \frac{p_{g} (X)I(G=g)}{1-p_{g} (X)  }\right]}   \right\} \{ Y_t - Y_{t-1} - m_{g, t} (X)  \}  \right] \label{eq: dr}
\end{align}
where $g\in \{a,b\}$, $p_{g} (X) = P(G=trt\mid X, G\in \{ trt, g\})$, $m_{g, t} (X) = E[ Y_{t} - Y_{t-1} \mid X, G = g ]$. All DID parameters in \eqref{eq: ipw}-\eqref{eq: dr} satisfy $\tau_{t,\rm ipw} (g)  = \tau_{t,\rm or} (g) =\tau_{t,\rm dr} (g) = \tau_{t,\rm adj} (g) $, where 
\begin{align}
	\tau_{t,\rm adj} (g)   & = E [Y_t - Y_{t-1}\mid G=trt] -  E\left[ \Delta_t (g, X) \mid G=trt\right] .  \label{eq: tau adj}
\end{align}

When $t=2$, \eqref{eq: ipw}-\eqref{eq: dr} are standard DID parameters calculated using each one of the two control groups, and 
\begin{align*}
	\tau_{2,\rm adj} (a)   &= ATT_2 + E\left[ \Delta_2 (trt, X)  -\Delta_2 (a, X) \mid G=trt\right]  \\
	\tau_{2,\rm adj} (b)   &= ATT_2 + E\left[ \Delta_2 (trt, X)  -\Delta_2 (b, X) \mid G=trt\right]  . 
\end{align*}
When $t>2$, \eqref{eq: ipw}-\eqref{eq: dr} are no longer standard DID parameters because $t-1$ is also a post-treatment period and 
\begin{align*}
	\tau_{t,\rm adj} (a)   &= ATT_t- ATT_{t-1} + E\left[ \Delta_t(trt, X)  -\Delta_t (a, X) \mid G=trt\right]  \\
	\tau_{t,\rm adj} (b)   &=  ATT_t- ATT_{t-1}  + E\left[ \Delta_t (trt, X)  -\Delta_t (b, X) \mid G=trt\right]  . 
\end{align*}
Under Assumption \ref{assump: monotone trends, X}, it is true that when $t=2$, the two DID parameters,  $	\tau_{2,\rm adj} (a) $ and $\tau_{2,\rm adj} (b) $, bound the $ATT_2$, i.e., $\min \{ 	\tau_{2,\rm adj} (a) , 	\tau_{2,\rm adj} (b)   \}\leq ATT_2 \leq  \max \{ 	\tau_{2,\rm adj} (a) , 	\tau_{2,\rm adj} (b)   \} $. When $t>2$, we have that $\min \{ 	\tau_{t,\rm adj} (a) , 	\tau_{t,\rm adj} (b)   \}\leq ATT_t-ATT_{t-1}  \leq  \max \{ 	\tau_{t,\rm adj} (a) , 	\tau_{t,\rm adj} (b)   \} $. 

The next theorem generalizes the partial  identification result in Theorem \ref{prop: 4} to the case that adjusts for observed covariates $X$.  

\begin{theorem}
	Under Assumption \ref{assump: monotone trends, X}, the average treatment effect for the treated $ATT_t, t= 2, \dots, T$ can be partially identified through
	\begin{eqnarray}
		&&\sum_{s=2}^{t}\min\{\tau_{s, \rm adj}(a), \tau_{s, \rm adj}(b)\} \leq ATT_t \leq \sum_{s=2}^{t}\max\{\tau_{s, \rm adj}(a), \tau_{s, \rm adj}(b)\},  \label{eq: new bound, X}
	\end{eqnarray}
	where $\tau_{t, \rm adj}(a)$ and $\tau_{t, \rm adj}(b)$ are defined in (\ref{eq: tau adj}). 
\end{theorem}

Results in this section can also directly apply to the case of staggered adoption  if we group treated states  according to their treatment adoption time, and construct two 
``never-treated'' control groups that satisfy the bracketing relationship.

\subsection{A data-driven way to construct control groups}

This method is based on the equivalent formulation in Lemma \ref{lemma: 1} and is data-driven. Suppose data from at least two time points in a ``prior-study'' period (i.e., $ t<1 $) are available, we identify two groups of control units whose relative outcomes compared with the treated group are negatively correlated during this prior-study period.

The steps are as follows. First, for each candidate control unit $ i $, let ${\bm \Gamma}(i)$ be the column vector of its average outcome relative to the treated group at every prior-study time period (i.e., the average outcome for every control unit after subtracting the average outcome for the treated group). Second, calculate the correlation matrix of $ {\bm \Gamma (i)} $'s. Third, use a hierarchical clustering algorithm to find two clusters of control units that exhibit strong between-cluster negative correlation. These two clusters of control units are then designated as control groups $ a,b $. The last two steps can be implemented in R using the \emph{corrplot} package \citep{corrplot}. By construction, control groups $ a,b $ satisfy Assumption \ref{assum4} in the prior-study period. We then must assume that this pattern persists during the study period. Conversely, if there are no control units which exhibit negative correlation, it indicates that Assumption \ref{assum4} may not be supported by the data and the proposed strategy for bracketing may not be applicable.

\subsection{Additional simulations on the bootstrap inference procedure}

In this section, we empirically evaluate the proposed modified bootstrap procedure (with $m=N/\log (\log N)$) and in a more challenging setting (Case III) with mild deviation from parallel trends. 

{\bf  Case III:  mild deviation from parallel trends:} $ E[Y^{(0)}_1 |G=trt] = 3, E[Y^{(0)}_1|G=a] = 10, E[Y^{(0)}_1|G=b] = 4$, $ \Delta_2 (trt)=1,  \Delta_3 (trt)=-4, \Delta_4 (trt)=1, \Delta_2 (a)=1,  \Delta_3 (a)=-4+\phi, \Delta_4 (a)=1,  \Delta_2 (b)=1-\phi,  \Delta_3 (b)=-4, \Delta_4 (b)=1+\phi  $, where $\phi=0.05, 0.01, 0.15,0.2,0.25$. 

In Case III, both $\sqrt{N} ( \theta_{\max}(P) - \theta_j(P) )$ and $\sqrt{N} ( \theta_j(P) - \theta_{\min}(P) )$ are bounded by 
$ \sqrt{N} ( \theta_{\max}(P) - \theta_{\min} (P) )$,  which is respectively $|\phi|, 2|\phi|, 3|\phi| $ for $ATT_2, ATT_3, ATT_4$. In other words, $\theta_{\min} (P)$ and $ \theta_{\max} (P)$ are not well-separated from the other parameters. This setting is to investigate the performance of the proposed methods when (S1) in Theorem \ref{theo} is violated.

\begin{table}[h]
	\caption{Simulation results based on 1000 runs. The statistics reported are:  simulation average of the half-median unbiased estimators $\hat\theta_{N, \min}^{\rm med}$  and  $\hat\theta_{N, \max}^{\rm med}$, the average CI length and coverage probability (CP in \%) of 95\% CIs obtained from the proposed bootstrap method in \eqref{eq: CI}-\eqref{eq: CI att} with either $m=N$ or  $m=N/\log (\log N)$,  and two other methods (the intersection-union method and  standard percentile bootstrap). The sample size is $ N=1000$ and number  of bootstrap iterations is $ B=300 $. 	\label{tb: simu S} } 
	\centering
	\fbox{	
		\resizebox{\textwidth}{!}{\begin{tabular}{lccccccccccccccccccc}  
				&  \multicolumn{6}{c}{Modified bootstrap  ($m=N$)}        &&      \multicolumn{6}{c}{ Modified bootstrap ($m=N/\log (\log N)$)}           && \multicolumn{2}{c}{Intersec.-Union} && \multicolumn{2}{c}{Perc. Boot.}\\
				&      $\hat\theta_{m, \min}^{\rm med}$ & $\hat\theta_{m, \max}^{\rm med}$            &      \multicolumn{2}{c}{CI  (Set) }           &      \multicolumn{2}{c}{CI  ($ ATT_t $)}     	&&        $\hat\theta_{m, \min}^{\rm med}$ & $\hat\theta_{m, \max}^{\rm med}$       &      \multicolumn{2}{c}{CI  (Set) }      &  \multicolumn{2}{c}{CI  ($ATT_t$) }        &&  \multicolumn{2}{c}{CI  (Set) }   &&  \multicolumn{2}{c}{CI  (Set) }    \\
				[0.5ex] 	\cline{2-7}\cline{9-14}  \cline{16-17} \cline{19-20}
				&	Mean& Mean &  Length & CP  & Length & CP 	&&Mean& Mean   &  Length     & CP & Length     & CP   &&  Length     & CP  &&  Length     & CP     \\  \hline 
				{\footnotesize \bf Case I  }   &                    &                    &           &       &      &         &           \\
				$ t=2 $ &1.970&2.030&0.483 & 96.7 & 0.478 & 96.7 && 1.966           & 2.039           & 0.654     & 96.3     & 0.649     & 96.2    && 0.553 & 99.0 &&0.581	&99.3\\
				$ t=3 $	&2.941 &	3.063 &0.583 & 97.7 & 0.575 & 97.7 &&   2.929           & 3.078           & 0.776     & 98.1     & 0.765     & 98.1 && 0.730 & 99.8 && 0.771	& 99.9\\
				$ t=4 $ &0.913	& 1.090 &	0.672 & 98.4 & 0.661 & 98.4 && 0.894           & 1.113           & 0.887     & 99.3     & 0.873     & 99.3  && 0.893 & 100.0 && 0.955	&100.0\\
				{\footnotesize \bf Case II    }    &                    &                    &           &       &      &         &           \\
				$ t=2 $	&1.003	& 1.997&1.455 & 97.9 & 1.404 & 96.8 && 1.005           & 2.000               & 1.636     & 96.9     & 1.575     & 96.2     && 1.443 & 97.7 && 1.455&	97.8 \\
				$ t=3 $		&-0.994	& 2.998&4.562 & 98.1 & 4.472 & 96.1 &&  -0.996          & 3.004           & 4.792     & 98.5     & 4.667     & 96.0       &&4.547	&97.8&&	4.563	&98.1 \\
				$ t=4 $ &-3.021 &	1.025	& 4.633 &	98.5 &4.542	& 96.7  &&	 -3.030           & 1.038           & 4.879     & 98.4     & 4.753     & 96.8     && 4.693	&98.7	&&4.728 &	99.3\\
				\multicolumn{2}{l}{\footnotesize \bf Case III 	($\delta$=0.05)} 	\\
				$ t=2 $		& 1.990 & 2.060 & 0.497 & 96.2  & 0.492 & 96.2  &  & 1.986 & 2.068 & 0.667 & 96.6  & 0.660 & 96.6  &  & 0.562 & 98.7  &  & 0.589 & 99.0  \\
				$ t=3 $		& 2.931 & 3.073 & 0.602 & 97.9  & 0.593 & 97.9  &  & 2.920 & 3.087 & 0.794 & 98.4  & 0.782 & 98.4  &  & 0.738 & 99.8  &  & 0.779 & 99.9  \\
				$ t=4 $	 & 0.874 & 1.079 & 0.704 & 98.0  & 0.691 & 98.0  &  & 0.856 & 1.102 & 0.918 & 99.3  & 0.902 & 99.3  &  & 0.911 & 100.0 &  & 0.972 & 100.0 \\
				\multicolumn{2}{l}{\footnotesize \bf Case III 	($\delta$=0.1)} 	\\ 
				$ t=2 $	& 1.998 & 2.102 & 0.535 & 95.3  & 0.528 & 95.2  &  & 1.996 & 2.107 & 0.701 & 96.1  & 0.693 & 96.1  &  & 0.586 & 98.3  &  & 0.609 & 98.9  \\
				$ t=3 $		& 2.902 & 3.102 & 0.660 & 99.2  & 0.647 & 99.1  &  & 2.896 & 3.112 & 0.845 & 99.1  & 0.830 & 99.1  &  & 0.765 & 99.9  &  & 0.806 & 100.0 \\
				$ t=4 $	 & 0.804 & 1.099 & 0.801 & 99.1  & 0.783 & 99.1  &  & 0.792 & 1.117 & 1.006 & 99.4  & 0.985 & 99.4  &  & 0.968 & 100.0 &  & 1.023 & 100.0 \\
				\multicolumn{2}{l}{\footnotesize \bf Case III 	($\delta$=0.15)}      \\
				$ t=2 $	 & 2.000 & 2.149 & 0.587 & 95.1  & 0.577 & 95.0  &  & 2.001 & 2.153 & 0.749 & 95.9  & 0.738 & 95.9  &  & 0.620 & 98.3  &  & 0.641 & 98.7  \\
				$ t=3 $	& 2.858 & 3.146 & 0.747 & 99.9  & 0.728 & 99.8  &  & 2.857 & 3.151 & 0.923 & 99.7  & 0.903 & 99.7  &  & 0.812 & 100.0 &  & 0.851 & 100.0 \\
				$ t=4 $		& 0.713 & 1.141 & 0.945 & 99.5  & 0.919 & 99.5  &  & 0.707 & 1.151 & 1.139 & 99.8  & 1.109 & 99.8  &  & 1.061 & 100.0 &  & 1.107 & 100.0 \\
				\multicolumn{2}{l}{\footnotesize \bf Case III 	($\delta$=0.2)}      \\
				$ t=2 $	 & 2.000 & 2.199 & 0.643 & 94.9  & 0.630 & 94.9  &  & 2.002 & 2.201 & 0.804 & 95.3  & 0.791 & 95.3  &  & 0.661 & 97.8  &  & 0.680 & 98.2  \\
				$t=3$	& 2.808 & 3.196 & 0.846 & 100.0 & 0.822 & 100.0 &  & 2.811 & 3.196 & 1.017 & 99.9  & 0.990 & 99.9  &  & 0.881 & 100.0 &  & 0.913 & 100.0 \\
				$t=4$& 0.612 & 1.191 & 1.108 & 99.8  & 1.074 & 99.7  &  & 0.613 & 1.196 & 1.294 & 99.9  & 1.257 & 99.9  &  & 1.181 & 100.0 &  & 1.217 & 100.0 \\
				\multicolumn{2}{l}{\footnotesize \bf Case III 	($\delta$=0.25)}      \\
				$t=2$ & 1.999 & 2.251 & 0.701 & 95.9  & 0.685 & 95.6  &  & 2.002 & 2.252 & 0.863 & 95.6  & 0.846 & 95.5  &  & 0.707 & 97.6  &  & 0.724 & 98.1  \\
				$t=3$ 	& 2.756 & 3.248 & 0.951 & 100.0 & 0.921 & 100.0 &  & 2.761 & 3.246 & 1.117 & 100.0 & 1.085 & 100.0 &  & 0.966 & 100.0 &  & 0.991 & 100.0 \\
				$t=4$ 	& 0.508 & 1.245 & 1.278 & 100.0 & 1.236 & 100.0 &  & 0.512 & 1.246 & 1.463 & 99.9  & 1.416 & 99.9  &  & 1.317 & 100.0 &  & 1.345 & 100.0
	\end{tabular}}}
\end{table}

From Table \ref{tb: simu S}, we see that the proposed bootstrap method with $m=N$ still can ensure adequate  coverage probability  across all settings. The  proposed bootstrap method with $m=N/\log (\log N)$,  is theoretically attractive according to Theorem \ref{theo}, but tends to generate confidence intervals that are much wider than the counterpart with $m=N$ because of the subsampling step. Nonetheless, in some cases when the bounding parameters are close to each other (e.g., Case I and $t=4$), it can still outperform the intersection-union method and the standard percentile bootstrap. Based on these observations, we still recommend our bootstrap procedure with $m=N$ in practice.

\section{Technical Proofs}

\subsection{Proof of (\ref{eq: union bounds})}
We  prove that the  bounds in \eqref{eq: new bound}  and  (\ref{eq: union bounds}) are equivalent. We will only prove  the lower bounds for $ ATT_t $ in  (\ref{eq: new bound})  and (\ref{eq: union bounds}) are equal, i.e.,  $ \sum_{s=2}^{t}\min\{\tau_s(a), \tau_s(b)\}  =  \min_{g_s\in \{ a, b\}}\{ \sum_{s=2}^t \tau_s (g_s)   \} $, the upper bound can be proved similarly.  

First, it is easy to see that $ \sum_{s=2}^{t}\min\{\tau_s(a), \tau_s(b)\}  \geq  \min_{g_s\in \{ a, b\}}\{ \sum_{s=2}^t \tau_s (g_s)   \} $, because $  \sum_{s=2}^{t}\min\{\tau_s(a), \tau_s(b)\}\in \{ \sum_{s=2}^t \tau_s (g_s) :g_s\in \{ a, b\} \} $ and the right hand side is the minimum. To prove the other direction, because $\min\{\tau_s(a), \tau_s(b)\} \leq \tau_s (g_s) $ for every $ s $, we have that $ \sum_{s=2}^{t}\min\{\tau_s(a), \tau_s(b)\}  \leq \sum_{s=2}^t \tau_s (g_s) $, $ g_s\in \{ a, b\} $. Hence, $ \sum_{s=2}^{t}\min\{\tau_s(a), \tau_s(b)\} \leq  \min_{g_s\in \{ a, b\}}\{ \sum_{s=2}^t \tau_s (g_s)   \}$. This completes the proof.

\subsection{Proof of Theorem \ref{theo}}

\begin{proof}
	(a) Consider the lower bound.  Let $\theta_{\min} (P) = \min_j \theta_j (P) $ and 
	$ {\cal M}(P)=\{j\in [k]: \theta_j(P)=\theta_{\min} (P) \}$ index the $ \theta_j(P) $'s that are equal to the minimum. Let $\hat\theta_{N,\min}= \min_j \hat\theta_{Nj}, T_{Nj} (P) = \sqrt{N} (\hat\theta_{Nj} - \theta_j (P)), T_{Nj}^* = \sqrt{N}  (\hat\theta_{Nj}^* - \hat\theta_{Nj}) $.

	Then for $L_m(x)=P\left\{ \sqrt{m} (  \hat{\theta}_{m,\min} -\theta_{\min}(P) )\leq x\right\}$,% for any $ \epsilon>0 $ and $ x\in {\cal R} $, there exists an $ N_{02} $, s.t. for $ N> N_{02} $, 
	\begin{align*}
		&L_m(x)\\
		&=P\left\{ \sqrt{m} (  \hat{\theta}_{m,\min} -\theta_{\min}(P) )\leq x\right\}\\ 
		&= P\left\{ \sqrt{m}  \min_j (\hat{\theta}_{mj} -\theta_{\min}(P) )\leq x\right\} \nonumber\\
		&= 1-  P\left\{ \sqrt{m}  \min_j (\hat{\theta}_{mj} -\theta_{\min}(P) )> x\right\} \nonumber\\
		&= 1- P\left\{  \sqrt{m} (\hat{\theta}_{mj} - \theta_j(P) )>x, j \in {\cal M}(P) , ~  \sqrt{m} (\hat{\theta}_{mj} - \theta_{\min} (P) )>x, j\notin {\cal M} (P) \right\} \nonumber\\
		&=   1- P\left\{ T_{mj} (P) >x, j \in {\cal M}(P) , ~  \sqrt{m} (\hat{\theta}_{mj} - \theta_{\min} (P) )>x, j\notin {\cal M}(P) \right\} \nonumber\\
		&=  1-  P\left\{ T_{mj} (P) >x, j \in {\cal M} (P), ~  T_{mj} (P)>x -  \sqrt{m} ({\theta}_{j} (P) - \theta_{\min} (P) ) , j\notin {\cal M}(P) \right\} \nonumber .
	\end{align*}

	%For $ j\notin {\cal M} $,  we have that $ \theta_j- \theta_1\geq  \kappa $ according to the assumptions, and therefore 	$ \sqrt{N} (\hat{\theta}_j-\min_{j'} \hat{\theta}_{j'}) $ can be written as
	%	\begin{align}
		%		\sqrt{N} (\hat{\theta}_j-\min_{j'} \hat{\theta}_{j'})&=\sqrt{N} \{ (\hat{\theta}_j- \theta_j) +  ( \theta_j - \theta_1) + (  \theta_1 -\min_{j'} \hat{\theta}_{j'} )\} \nonumber\\
		%		&\geq  \sqrt{N}  (\hat{\theta}_j- \theta_j)  +  \sqrt{N} \kappa + \sqrt{N} ( \theta_1- \hat{\theta}_m) \qquad m\in {\cal M} \label{eq: app1}
		%	\end{align}
	%	where the second inequality is because $ \theta_j- \theta_0\geq \kappa $ for $ j \notin {\cal M} $ and $ \min_{j'} \hat{\theta}_{j'} \leq \hat{\theta}_m$. Then (\ref{eq: app1}) would be arbitrarily large because $ \sqrt{N} (\hat{\theta}_j- \theta_j)  $ and $  \sqrt{N} (\theta_0- \hat{\theta}_m)$ are both $ O_p(1) $ and $ \sqrt{N} \kappa $ goes to infinity. Specifically,  for any $\epsilon>0, C_\epsilon>0$, there  exists an $N_{01}$ s.t. for $N>N_{01}$,  $P\{\sqrt{N} (\hat{\theta}_j-\min_{j'} \hat{\theta}_{j'})\leq C_\epsilon\}\leq \epsilon$ when $ j\notin {\cal M} $. Therefore, for any $ \epsilon>0 $ and $ x\in {\cal R} $, there exists an $ N_{01} $, s.t. for $ N> N_{01} $, 
	
	We consider the proposed modified bootstrap procedure.	 Recall that $d_{j, \min} = (1- \sqrt{m}/\sqrt{N}) ( \hat\theta_{N,\min}  - \hat \theta_{Nj} )$ and thus $\sqrt{N} d_{j, \min}  + \sqrt{N} (\hat\theta_{Nj}  - \hat\theta_{N,\min})  =  \sqrt{m} (\hat\theta_{Nj} - \hat\theta_{N,\min} ) $. Also let $\hat\theta_{Nj, {\rm mod}}^*=\hat\theta_{Nj}^* + d_{j, \min}  $ and  $V_{Nj} = \hat\theta_{N, \min}  - \hat\theta_{Nj} - \theta_{\min} (P) + \theta_j (P)$ for $j\in [k]$, we have 	
	\begin{align*}
		& \hat{L}_{N,{\rm mod}}(x) \\
		&=P_*\left\{ \sqrt{N} (\min_j \hat{\theta}_{Nj, {\rm mod}}^* - \hat{\theta}_{N,\min})\leq x\right\}\\ 
		&= P_*\left\{ \sqrt{N}\min_j (\hat{\theta}_{Nj, {\rm mod}}^*- \hat{\theta}_{N,\min}) \leq x\right\}\nonumber \\
		&=1-  P_*\left\{ \sqrt{N}\min_j (\hat{\theta}_{Nj, {\rm mod}}^*- \hat{\theta}_{N,\min}) > x\right\}\nonumber \\
		&=  1-  P_*\left\{ \sqrt{N} (\hat\theta_{Nj}^* + d_{j, \min}  - \hat{\theta}_{N,\min} )  > x, ~\forall j \right\}\nonumber \\
		&=  1-  P_*\left\{ \sqrt{N} (\hat\theta_{Nj}^* - \hat\theta_{Nj}+ d_{j, \min} + \hat\theta_{Nj}  - \hat{\theta}_{N,\min} )  > x, ~\forall j \right\}\nonumber \\
		&=  1-  P_*\left\{ \sqrt{N} (\hat\theta_{Nj}^* - \hat\theta_{Nj})  > x -\sqrt{N} ( d_{j, \min}  + \hat\theta_{Nj} -\hat\theta_{N,\min} ), ~\forall j \right\}\nonumber \\
		&=  1-  P_*\left\{ \sqrt{N} (\hat\theta_{Nj}^* - \hat\theta_{Nj})  > x + \sqrt{m} ( \hat\theta_{N,\min} - \hat\theta_{Nj}  ), ~\forall j \right\}\nonumber \\
		&=   1-  P_*\left\{ \sqrt{N} (\hat\theta_{Nj}^* - \hat\theta_{Nj})  > x + \sqrt{m}V_{Nj} , ~j \in \mathcal{M} (P) \right. \\
		& \quad\quad\quad\quad \quad \quad  \left.   \sqrt{N} (\hat\theta_{Nj}^* - \hat\theta_{Nj})  > x + \sqrt{m}V_{Nj} + \sqrt{m} (\theta_{\min}(P) - \theta_j (P)), ~ j \notin \mathcal{M} (P)  \right\}\nonumber \\ 
		&=   1-  P_*\left\{ T_{Nj}^*  > x + \sqrt{m}V_{Nj} , j \in \mathcal{M} (P),  ~  T_{Nj}^*   > x + \sqrt{m}V_{Nj} - \sqrt{m} (\theta_j (P)- \theta_{\min}(P) ),  j \notin \mathcal{M} (P)  \right\}\nonumber \\ 
		&\leq 1- P_*\left\{  T_{Nj}^*  > x, j \in {\cal M}(P), ~ T_{Nj}^*  -   \sqrt{m}V_{Nj}  > x - \sqrt{m} (\theta_j (P)- \theta_{\min}(P) ) , j \notin {\cal M} (P) \right\} \nonumber\\
		&: =  \tilde {L}_{N,{\rm mod}}(x), 
	\end{align*}
	where the last inequality is because $V_{Nj} \leq 0  $ for $j \in \mathcal{M}(P)$.

	We prove the result under (S2) first.  Define 
	\begin{align*}
		\tilde 	L_N(x) &= 1-  P\left\{ T_{Nj} (P) >x, j \in {\cal M} (P), T_{Nj} (P)>x -  \sqrt{m} ({\theta}_{j} (P) - \theta_{\min} (P) ) , j\notin {\cal M}(P) \right\}  \\
		\tilde{L}^*_{N,{\rm mod}}(x)  & = 1- P_*\left\{  T_{Nj}^*  > x, j \in {\cal M}(P), ~ T_{Nj}^*   > x - \sqrt{m} (\theta_j (P)- \theta_{\min}(P) ) , j \notin {\cal M} (P) \right\} 
	\end{align*}
	
	By Lemma 11.4.2 of \cite{lehmann2005testing}, for any sequence $\{ P_N\in \mathcal{P}, N\geq 1 \}$, 
	we have that for every $ g $ and $ t $, 
	$$\frac1N \sum_{G_i =g} R_{it}  \xrightarrow{P_N} P(G_i = g, R_{it} = 1 ). $$
	Then by the uniform integrability of $ r_{igt} $ and 
	Lemma 11.4.1 of \cite{lehmann2005testing}, 	 and from  the proof of Theorem 3.7 in  \cite{Romano:2012aa}, we know that uniformly over $ P\in \mathcal{P} $,   $ \sqrt{N} ( \bar{\bm X}_N- \bm \mu (P) )  $ under $ P $ and  $ \sqrt{N} ( \bar{\bm X}^*- \bar{\bm X}_N )  $ under $  P_* $ converge to the same multivariate normal distribution. For $j=1,\dots, k$,  because $\theta_j (P) = {\bm c}_j^T \bm \mu (P)$ with $\bm c_j$ a  vector of fixed constants, we have   by the Cramer-Wold device that 
	\[
	\sup_{P\in \mathcal{P}} \sup_{x\in \mathcal{R}} | \tilde 	L_N(x)  -  \tilde L_{N, \rm mod}^* (x) | \rightarrow 0. 
	\]
	From Theorem 3.1 of \cite{Romano:2012aa}, as $m/N\rightarrow 0, m\rightarrow\infty $, we have that 
	\[
	\sup_{P\in \mathcal{P}} \sup_{x\in \mathcal{R}} |\tilde 	L_N(x)  - L_m (x)  | \rightarrow 0. 
	\]
	Hence, by triangle  inequality, we have 
	\[
	\sup_{P\in \mathcal{P}} \sup_{x\in \mathcal{R}} | \tilde L_{N, \rm mod}^* (x) - L_m (x)   | \rightarrow 0. 
	\]
	Moreover, as $m/N\rightarrow 0$, we have $\sqrt{m}V_{Nj} \xrightarrow{P_N} 0 $ for any $P_N \in \mathcal{P}$, and thus \\ $\sup_{P\in \mathcal{P}} \sup_{x\in \mathcal{R}} | \tilde L_{N, \rm mod} (x) - L_m (x)   | \rightarrow 0. $ Since we showed that $\hat L_{N,\rm mod} (x) \leq \tilde L_{N,\rm mod} (x)$, we conclude that  $\lim_{N\rightarrow\infty} \sup_{P\in \mathcal{P}} \sup_{x\in {\cal R}} \{\hat L_{N,\rm mod} (x)-L_m(x)\}\leq 0$. The upper bound can be proved in the same way and is omitted.

	Now consider the result under (S1) where $m=N$ and $\lim_{N\rightarrow \infty}  \inf_{P\in \mathcal{P}}  \min_{j \notin \mathcal{M}(P)} \sqrt{N} (\theta_j (P)-  \theta_{\min} (P)) = \infty $. Under this case, the modification term $d_j= 0$ for all $j$. In the definitions of $L_m(x)$ and $\tilde L_{N, \rm mod} (x)$, 
	the events involving $j\notin \mathcal{M}(P)$ hold with probability approaching 1 in a uniform sense. Hence we have $\sup_{P\in \mathcal{P}} \sup_{x\in \mathcal{R}} | \tilde L_{N, \rm mod} (x) - L_m (x)   | \rightarrow 0 $, and  thus  $\lim_{N\rightarrow\infty} \sup_{P\in \mathcal{P}} \sup_{x\in {\cal R}} \{\hat L_{N,\rm mod} (x)-L_m(x)\}\leq 0$.

	%	In the other scenario when $\theta_1=\theta_2$, 
	%	\begin{align}
		%	H_n(x) &=  P\left\{ \sqrt{n} \min (\hat{\theta}_1-\theta_1, \hat{\theta}_2-\theta_2) \leq x\right\}\nonumber\\
		%	&= 1- P\left\{ \sqrt{n} (\hat{\theta}_1-\theta_1) >x, \sqrt{n} (\hat{\theta}_2-\theta_2) >x\right\} \nonumber
		%	\end{align}

	(b) From the definition of $c_L^*(p)$, it satisfies $\hat{L}_{N, \rm mod}(c_L^*(p))\geq p$. From part (a), for any $\epsilon>0$, there exists an $N_{01}$ s.t. for $N>N_{01}$, $\sup_{P\in \mathcal{P}}\sup_{x\in \mathcal{R}} \{ \hat{L}_{N, \rm mod}(x)-L_m(x)\}\leq \epsilon$, and for every $ P\in \mathcal{P} $, 
	\begin{align}
		L_m(c_L^*(p))&\geq \hat{L}_{N, \rm mod}(c_L^*(p)) -\sup_{P\in \mathcal{P}} \sup_{x\in \mathcal{R}} \{\hat{L}_{N, \rm mod}(x) -L_m(x)\} \nonumber\\
		& \geq p -\epsilon \nonumber
	\end{align}
	This completes the proof of the first part in (b). 
	
	From the definition of $ c_U^*(1-p) $, it satisfies $ \hat{R}_{N, \rm mod}(c_U^*(1-p)) \leq 1-p$. From part (a), for any $ \epsilon>0 $, there exists an $ N_{02} $ s.t. for $ N>N_{02} $,  $\inf_{P\in \mathcal{P}}\inf_{x\in \mathcal{R}} \{ \hat{R}_{N, \rm mod}(x)-R_m(x)\}\geq -\epsilon$, and for every $P\in \mathcal{P} $,
	\begin{align}
		R_m(c_U^*(1-p))&\leq \hat{R}_{N, \rm mod}(c_U^*(1-p))  - \inf_{P\in \mathcal{P}} \inf_{x\in \mathcal{R}} \{\hat{R}_{N, \rm mod}(x) -R_m(x)\} \nonumber\\
		& \leq 1-p +\epsilon \nonumber
	\end{align}
	Therefore, for every $P\in \mathcal{P} $,
	\begin{align}
		& P\left\{ \sqrt{m}( \hat{\theta}_{m, \max} - \theta_{\max}(P))\geq c_U^* (1-p)\right\}\geq  1-  R_m( c_U^*(1-p))\geq p-\epsilon \nonumber
	\end{align}
	This completes the proof of (b).

	(c) Let $ p=1-\alpha/2 $ and rearrange,
	\begin{align}
		&\lim_{N\rightarrow \infty} \inf_{P\in \mathcal{P}} P\left\{  \hat{\theta}_{m, \min} \leq \theta_{\min}(P)+m^{-1/2}c_L^* (1-\alpha/2)\right\}\geq 1-\alpha/2 \nonumber\\
		&\lim_{N\rightarrow \infty} \inf_{P\in \mathcal{P}} P\left\{  \hat{\theta}_{m, \max}\geq \theta_{\max} +m^{-1/2}c_U^* (\alpha/2)\right\}\geq 1-\alpha/2 \nonumber
	\end{align}
	By Bonferroni's inequality, we have for every $ P\in \mathcal{P} $, 
	\begin{align}
		&P\left\{ [ \theta_{\min}(P),  \theta_{\max}(P)] \in CI_{1-\alpha} \right\} \nonumber\\
		&\geq 1- P\left\{  \theta_{\min}(P)  < \hat{\theta}_{m,\min}- m^{-1/2} c_L^*(1-\alpha/2)\right\} -P\left\{ \theta_{\max} (P)>\hat{\theta}_{m,\max}- m^{-1/2} c_U^* (\alpha/2)\right\} \nonumber\\
		& = P\left\{  \theta_{\min} (P) \geq \hat{\theta}_{m,\min}- m^{-1/2} c_L^*(1-\alpha/2)\right\} + P\left\{ \theta_{\max} (P)\leq  \hat{\theta}_{m,\max}- m^{-1/2} c_U^*(\alpha/2)\right\} -1 \nonumber
	\end{align}
	Therefore, 
	\begin{align}
		\lim_{n\rightarrow\infty} \inf_{P\in \mathcal{P} } P\left\{  [ \theta_{\min}(P),  \theta_{\max}(P)] \in CI_{1-\alpha}  \right\}\geq 1-\alpha/2 +1-\alpha/2-1=1-\alpha \nonumber
	\end{align}
	
	(d)   Define  $ p^\psi (P)= 1-\Phi\{\rho (\theta_{\max} (P) -\theta_{\min} (P))\}\alpha$.  From the condition, we have that {$ \hat{p}=p^\psi (P_N)+o_{P_N}(1) $ for any $P_N\in \mathcal{P}$}. 
	
	Decompose the probability that $ \psi_0 (P) $ is outside $ CI_{1-\alpha}^\psi $ as 
	\begin{align}
		P\left\{ \psi_0 (P)  \notin  CI_{1-\alpha}^\psi\right\}  & \leq \underbrace{P \left\{  \psi_0 (P) < \hat\theta_{m,\min} - m^{-1/2} c_L^*(\hat{p}) \right\}}_{A_L} \nonumber \\
		&\qquad + \underbrace{P \left\{  \psi_0 (P)  >  \hat\theta_{m,\max} - m^{-1/2} c_U^*(1-\hat{p}) \right\}}_{A_U} . \nonumber
	\end{align}
	Because with probability approaching 1, we have $ \hat{p} \geq p^\psi (P) -\epsilon/2$, and thus $ c_L^* (\hat{p}) \geq c_L^* (p^\psi (P)-\epsilon/2)  $  and $ c_U^* (1-\hat{p}) \leq c_U^* (1- p^\psi (P)+\epsilon/2)  $ because $ c_L^* (p)$ and $ c_U^* (p)$ are both non-decreasing functions of $ p $. Hence, the first component satisfies
	\begin{align}
		A_L & = P \left\{  \psi_0 (P) +m^{-1/2}  c_L^*(\hat{p}) < \hat\theta_{m,\min} \right\}\nonumber \\
		&\leq \underbrace{P \left\{  \psi_0 (P) +m^{-1/2}  c_L^* (p^\psi (P)-\epsilon/2) < \hat\theta_{m,\min} \right\}}_{  \tilde{A}_L  } +o(1)\nonumber
	\end{align}
	the second component satisfies 
	\begin{align}
		A_U & = P \left\{  \psi_0 (P) +m^{-1/2}  c_U^*(1-\hat{p}) > \hat\theta_{m,\max} \right\}\nonumber\\
		&\leq \underbrace{P \left\{  \psi_0 (P)+m^{-1/2}   c_U^* (1- p^\psi (P)+\epsilon/2)> \hat\theta_{m,\max} \right\}}_{ \tilde{A}_U} +o(1) \nonumber
	\end{align}
	In the following, we will show that for any $ \epsilon>0 $, there exists an $ N_{0}  $, s.t.  for $ N>N_{0} $, $ \tilde{A}_L+\tilde{A}_U\leq \alpha+\epsilon $. Define $ 
	\lambda $ as the limit:  $ \rho ( \theta_{\max} (P) - \theta_{\min} (P) ) \rightarrow \lambda \in [0, \infty]$.
	
	Suppose first $ \lambda =0 $ and $ p^\psi (P)=1-\alpha/2+o(1) $. For the same $ \epsilon $, there exists an $ N_{04} $, s.t. for $ N>N_{04} $, $ p^\psi (P)\geq 1-\alpha/2-\epsilon /2 $. 
	In this case, 
	\begin{align}
		\tilde{A}_L\leq P \left\{ \theta_{\min}(P) +m^{-1/2}  c_L^* (p^\psi (P)-\epsilon/2) <  \hat{\theta}_{m,\min} \right\} \leq \alpha/2+\epsilon \nonumber
	\end{align}
	where the first inequality is because $ \psi_0 (P)\geq \theta_{\min}(P) $, the second inequality uses $ p^\psi (P) \geq 1-\alpha/2-\epsilon/2  $ and Theorem \ref{theo}(b). Similarly, 
	\begin{align}
		\tilde{A}_U\leq P \left\{   \theta_{\max}(P) +m^{-1/2}  c_U^* (1-p^\psi (P)+\epsilon/2) >  \hat{\theta}_{m,\max} \right\} \leq \alpha/2+\epsilon \nonumber
	\end{align}
	Hence, when $ N>  N_{04} $,  $ \lambda =0 $, we have $ \tilde{A}_L+\tilde{A}_U\leq \alpha+2\epsilon $.
	
	Then, consider $ \lambda \in (0, \infty] $ and $ p^\psi (P)= 1-\Phi(\lambda ) \alpha+o(1)$. For the same $ \epsilon $, there exists an $ N_{05} $, s.t. for $ N> N_{05} $, $ p^\psi (P)\geq 1- \Phi(\lambda )\alpha-\epsilon/2 $.  Also since $ \lambda >0 $, $ m^{1/2} ( \theta_{\max} (P)-  \theta_{\min} (P) ) \rightarrow \infty$. Without loss of generality, assume $ m^{1/2} (\psi_0 (P)- \theta_{\min}(P) ) \rightarrow \infty$. Under this circumstance, 
	\begin{align}
		\tilde{A}_L&= P\left\{  \sqrt{m} ( \hat{\theta}_{m,\min} -  \theta_{\min}(P) )> c_L^* (p^\psi (P)-\epsilon/2) + m^{1/2} (\psi_0 (P)- \theta_{\min} (P)) \right\}\nonumber\\
		&\leq  P\left\{  \sqrt{m} (\hat{\theta}_{mj} - \theta_j (P) )> c_L^* (p^\psi (P)-\epsilon/2) + m^{1/2} (\psi_0 (P)- \theta_{\min} (P)) \right\} \qquad (j \in \mathcal{M} (P)  ) \nonumber\\
		&=o(1) \nonumber\\
		\tilde{A}_U& = P \left\{ \psi_0 (P)- \theta_{\max}(P) +m^{-1/2}  c_U^* (1-p^\psi (P)+\epsilon/2) >  \hat{\theta}_{m,\max} - \theta_{\max} (P) \right\}\nonumber\\
		&\leq P \left\{  m^{1/2} ( \hat{\theta}_{m,\max} - \theta_{\max} (P))< c_U^* (1-p^\psi (P)+\epsilon/2) \right\}\nonumber\\
		&\leq P\left\{ m^{1/2} ( \hat{\theta}_{m,\max} - \theta_{\max} (P)) <c_U^* \left( \Phi(\lambda ) \alpha +\epsilon\right)   \right\} \leq \Phi(\lambda ) \alpha+\epsilon \nonumber
	\end{align}
	For the same $ \epsilon $, there exists an $ N_{06} $, when $ N> N_{06} $, $ \tilde{A}_L \leq \epsilon$.  Hence, when $ N> \max ( N_{05}, N_{06}) $, $ \tilde{A}_L+\tilde{A}_U\leq \Phi(\lambda )\alpha+2\epsilon $.

	Combined, we have for any $ \epsilon>0 $, for $ N> \max(N_{04}, N_{05}, N_{06}) $, $ P\left\{ \psi_0 \notin  CI_{1-\alpha}^\psi\right\} \leq A_L+A_U \leq \alpha+2\epsilon$, which completes the proof for Theorem \ref{theo}(d). 
\end{proof}

\subsection{Proof of the Uniform Validity of Berger \& Boos (1994)} 
\begin{theorem}
	For $ t=1,\dots, T $ and $ g=a, b, trt $, 
	suppose that 	$r_{igt}=  \{Y_{it}- E(Y_{it}\mid G_i=g)\} I(G_i= g) R_{it} / P(G_i=g, R_{it}=1)$ is uniformly integrable in the sense that
	\[
	\lim_{\lambda\rightarrow \infty} \sup_{P\in \mathcal{P}} E_P\left\{  \frac{r_{igt}^2}{\var (r_{igt}) }  I\left( \frac{|r_{igt}|}{\var (r_{igt})^{1/2}} >\lambda\right) \right\}  =0,
	\]
	$ \theta_j(\bm\mu) = \bm c_j^T \bm\mu $ with $\bm c_j $ a vector of fixed constants for $ j =1,\dots, k$,  then  
	\begin{align}
		CI_{1-\alpha}^{\rm union}\equiv \left[  \min_j ( \hat\theta_{Nj} - z_{1-\alpha/2} \hat\sigma_{Nj}),   \max_j ( \hat\theta_{Nj} + z_{1-\alpha/2} \hat\sigma_{Nj})\right] \label{eq: CI union}
	\end{align} is a uniformly valid $1-\alpha$ level CI for  the identified set $\Psi_0 (P)=[ \theta_{\min}(P), \theta_{\max}(P) ]$, i.e., $ 	\lim_{N\rightarrow\infty} \inf_{P\in{\cal P}} P\big( \Psi_0(P) \subseteq CI_{1-\alpha}^{\rm  union} \big) \geq 1-\alpha  $.\\
\end{theorem}
\begin{proof}
	By the uniform integrability of $r_{igt}$ and Lemma 11.4.1 of \cite{lehmann2005testing}, we have that for any sequence $\{ P_N \in \mathcal{P}, N\geq 1\}$ and any $j$, 
	\[
	\frac{	 \hat\theta_{Nj} - \theta_j(P) }{\hat\sigma_{Nj}} \xrightarrow{d} N(0,1). 
	\]
	Hence $\lim_{N\rightarrow\infty} \inf_{P\in{\cal P}} P\big(  \theta_j (P) \leq \hat\theta_{Nj}+ z_{1-\alpha/2} \hat\sigma_{Nj}  \big) = 1-\alpha/2 $ for every $j$. Hence 
	\begin{align*}
		&	\lim_{N\rightarrow\infty} \inf_{P\in{\cal P}} P\big(   \theta_{\max} (P) \leq  \max_j (\hat\theta_{Nj}+ z_{1-\alpha/2} \hat\sigma_{Nj} ) \big)  \\
		&= \lim_{N\rightarrow\infty} \inf_{P\in{\cal P}} P\big(   \theta_j (P) \leq  \max_j (\hat\theta_{Nj}+ z_{1-\alpha/2} \hat\sigma_{Nj} ) \big)  \quad \text{for some $j$ such that $\theta_j (P)= \theta_{\max} (P) $}\\
		&\geq  \lim_{N\rightarrow\infty} \inf_{P\in{\cal P}} P\big(   \theta_j (P) \leq   \hat\theta_{Nj}+ z_{1-\alpha/2} \hat\sigma_{Nj}  \big)  \\
		&= 1- \alpha/2. 
	\end{align*}

	Similarly, we have that $	\lim_{N\rightarrow\infty} \inf_{P\in{\cal P}} P\big(   \theta_{\min} (P) \geq  \min_j (\hat\theta_{Nj} - z_{1-\alpha/2} \hat\sigma_{Nj} ) \big) =1-\alpha/2$. Then by Bonferroni's inequality, we have proved the conclusion that $ 	\lim_{N\rightarrow\infty} \inf_{P\in{\cal P}} P\big( \Psi_0(P) \subseteq CI_{1-\alpha}^{\rm  union} \big) \geq 1-\alpha  $. 
\end{proof}

\subsection{Proof of Theorem \ref{prop: SA}}
(a)  Define $ ATT_1 =0$. From (\ref{eq: tau})-(\ref{eq: tau t}), we have for every $ s\geq2 $,
\begin{align}
	&\min(\tau_s(a), \tau_s(b))= ATT_s-ATT_{s-1} + \Delta_s^{(0)} (trt)- \max( \Delta_s^{(0)}(a), \Delta_s^{(0)}(b))  \nonumber\\
	&\max(\tau_s(a), \tau_s(b))=ATT_s-ATT_{s-1} + \Delta_s^{(0)} (trt)- \min( \Delta_s^{(0)}(a), \Delta_s^{(0)}(b))   \nonumber
\end{align}
Then, from Assumption \ref{asump: 5}, 
\begin{align}
	&\min(\tau_s(a), \tau_s(b))\leq  ATT_s-ATT_{s-1} +\delta_s  \nonumber\\
	&\max(\tau_s(a), \tau_s(b))\geq ATT_s-ATT_{s-1} -\gamma_s  \nonumber
\end{align}
Hence, 
\begin{eqnarray}
	\min(\tau_2 (a), \tau_2(b))-\delta_2 \leq & ATT_2 &\leq \max(\tau_2 (a), \tau_2(b))+ \gamma_2\nonumber\\
	\min(\tau_2 (a), \tau_2(b))-\delta_s \leq & ATT_s-ATT_{s-1} &\leq \max(\tau_2 (a), \tau_2(b))+ \gamma_s\nonumber
\end{eqnarray}
Theorem \ref{prop: SA}(a) is proved by summing these inequalities. \\
(b) Let $ [\hat{l}_t, \hat{r}_t] $ be the  confidence interval for the identified set under Assumption \ref{assum4}. Then, 
\begin{align}
	&P\left(\left[\sum_{s=2}^{t}\min\{\tau_s(a), \tau_s(b)\} , \sum_{s=2}^{t}\max\{\tau_s(a), \tau_s(b)\} \right]\in [\hat{l}_t, \hat{r}_t]\right) \nonumber  \\
	= & P\left(\sum_{s=2}^{t}\min\{\tau_s(a), \tau_s(b)\} \geq \hat{l}_t,  \sum_{s=2}^{t}\max\{\tau_s(a), \tau_s(b)\}\leq \hat{r}_t\right) \nonumber \\
	= &P\left(\sum_{s=2}^{t}\min\{\tau_s(a), \tau_s(b)\} - \sum_{s=2}^t \delta_s \geq \hat{l}_t - \sum_{s=2}^t \delta_s ,  \sum_{s=2}^{t}\max\{\tau_s(a), \tau_s(b)\} + \sum_{s=2}^{t}\gamma_s\leq \hat{r}_t+ \sum_{s=2}^{t}\gamma_s\right) \nonumber \\
	=&P\left(\left[\sum_{s=2}^{t}\min\{\tau_s(a), \tau_s(b)\} - \sum_{s=2}^t \delta_s , \sum_{s=2}^{t}\max\{\tau_s(a), \tau_s(b)\}+ \sum_{s=2}^t \gamma_s \right]\in [\hat{l}_t - \sum_{s=2}^t \delta_s, \hat{r}_t+ \sum_{s=2}^t \gamma_s ]\right) \nonumber  
\end{align}
The uniform validity of $ [\hat{l}_t- \sum_{s=2}^t \delta_s, \hat{r}_t+ \sum_{s=2}^t \gamma_s] $ is directly from the uniform validity of $ [\hat{l}_t, \hat{r}_t] $.

Next, we prove the results hold for the parameter of interest $ ATT_t $, where $ ATT_t $ lies in the identified set (\ref{eq: bound SA}). This proof is based on and is similar to the proof of Theorem \ref{theo}(d), and thus some details are omitted. 
Let $ [\hat{l}_t, \hat{r}_t] $ be the confidence interval for $ ATT_t $ under Assumption \ref{assum4}. Under Assumption \ref{asump: 5}, decompose the probability that $ ATT_t $ is outside $ [\hat{l}_t- \sum_{s=2}^t \delta_s, \hat{r}_t- \sum_{s=2}^t \gamma_s] $ as
\begin{align}
	&	P\left( ATT_t\notin  \left[\hat{l}_t- \sum_{s=2}^t \delta_s, \hat{r}_t- \sum_{s=2}^t \gamma_s\right] \right) \nonumber\\
	& \leq P\left(  ATT_t <\hat{l}_t- \sum_{s=2}^t \delta_s \right)+ P\left(  ATT_t >\hat{r}_t+ \sum_{s=2}^t \gamma_s \right) \nonumber
\end{align}
Consider the two scenarios when constructing $ [\hat{l}_t, \hat{r}_t] $ as in the proof of Theorem \ref{theo}(d). Define $ \lambda $ still as the limit based on the identified set in (\ref{eq: new bound}),
\[
\rho\left[ \sum_{s=2}^t \max \{ \tau_s(a), \tau_s(b)\} - \sum_{s=2}^t \min \{ \tau_s(a), \tau_s(b)\} \right] \rightarrow \lambda\in[0, \infty].
\]

First, consider $ \lambda=0 $, that is when $ p^\psi=1-\alpha/2+o(1) $ in constructing $ [\hat{l}_t, \hat{r}_t] $. In this scenario,
\begin{align}
	&P\left(  ATT_t <\hat{l}_t- \sum_{s=2}^t \delta_s \right)\leq P\left(  \sum_{s=2}^t \min \left\{ \tau_s (a), \tau_s(b)\right\} - \sum_{s=2}^t \delta_s<\hat{l}_t- \sum_{s=2}^t \delta_s \right) \nonumber\\
	&=  P\left(  \sum_{s=2}^t \min \left\{ \tau_s (a), \tau_s(b)\right\} <\hat{l}_t\right) \nonumber \leq \alpha/2+o(1)
\end{align}
Similarly, 
\begin{align}
	&P\left(  ATT_t >\hat{r}_t+\sum_{s=2}^t \gamma_s \right)\leq\alpha/2+o(1) \nonumber
\end{align}

Consider the second scenario when $ \lambda\in(0, \infty] $ and $ p^\psi=1-\Phi(\lambda)\alpha+ o(1) $ in constructing $ [\hat{l}_t, \hat{r}_t] $. Without loss of generality, assume $ N^{1/2} [ATT_t -\{ \sum_{s=2}^t \min \left\{ \tau_s (a), \tau_s(b)\right\} \}]\rightarrow \infty $, and thus, $ N^{1/2} [ATT_t+\sum_{s=2}^t \delta_s -\{ \sum_{s=2}^t \min \left\{ \tau_s (a), \tau_s(b)\right\} \}]\rightarrow \infty $. Under this circumstance, 
\begin{align}
	&P\left(  ATT_t <\hat{l}_t- \sum_{s=2}^t \delta_s \right)\leq P\left(  ATT_t+\sum_{s=2}^t \delta_s <\hat{l}_t \right) =o(1)\nonumber\\
	&P\left(  ATT_t >\hat{r}_t+ \sum_{s=2}^t \gamma_s \right) \leq P\left(  \sum_{s=2}^t \max \left\{ \tau_s (a), \tau_s(b)\right\} +\sum_{s=2}^t \gamma_s>\hat{r}_t+ \sum_{s=2}^t \gamma_s \right) \nonumber\\
	=&  P\left(  \sum_{s=2}^t \max \left\{ \tau_s (a), \tau_s(b)\right\} >\hat{r}_t\right) \leq \Phi(\lambda)\alpha+o(1)\nonumber
\end{align}
Combining two scenarios and using similar arguments as in the proof of theorem \ref{theo}(d), uniform validity is established, that is 
\begin{align}
	\lim_{N\rightarrow \infty} \inf_{P\in\mathcal{P}} \inf_{ATT_t}P\left( ATT_t\notin  \left[\hat{l}_t- \sum_{s=2}^t \delta_s, \hat{r}_t+ \sum_{s=2}^t \gamma_s\right] \right) \leq \alpha \nonumber
\end{align}

\end{document}